\documentclass[reprint,onecolumn,superscriptaddress]{revtex4-2}
\usepackage[utf8]{inputenc}
\usepackage{graphicx}
\usepackage{hyperref}
\usepackage{amsmath}
\usepackage{amssymb}
\usepackage{tabularx}
\usepackage{booktabs}
\usepackage{lineno}

\usepackage{bm}

\hypersetup{
    colorlinks=true,
    linkcolor=blue,
    filecolor=blue,
    urlcolor=blue,
    citecolor=blue,
}

\begin{document}
\title{Topological superconductor candidates PdBi$_2$Te$_4$ and PdBi$_2$Te$_5$ from a generic ab initio strategy}

\author{Aiyun Luo}
\thanks{These authors made equal contributions to this work.}
\affiliation{Wuhan National High Magnetic Field Center $\&$ School of Physics, Huazhong University of Science and Technology, Wuhan 430074, China}
\author{Ying Li}
\thanks{These authors made equal contributions to this work.}
\affiliation{Wuhan National High Magnetic Field Center $\&$ School of Physics, Huazhong University of Science and Technology, Wuhan 430074, China}
\author{Yi Qin}
\thanks{These authors made equal contributions to this work.}
\affiliation{Wuhan National High Magnetic Field Center $\&$ School of Physics, Huazhong University of Science and Technology, Wuhan 430074, China}
\author{Jingnan Hu}
\affiliation{Wuhan National High Magnetic Field Center $\&$ School of Physics, Huazhong University of Science and Technology, Wuhan 430074, China}
\author{Biao Lian}
\affiliation{Department of Physics, Princeton University, Princeton, NJ 08544, United States of America}
\author{Gang Xu}
\email[e-mail address: ]{gangxu@hust.edu.cn}
\affiliation{Wuhan National High Magnetic Field Center $\&$ School of Physics, Huazhong University of Science and Technology, Wuhan 430074, China}
\affiliation{Institute for Quantum Science and Engineering, Huazhong University of Science and Technology, Wuhan 430074, China}
\affiliation{Wuhan Institute of Quantum Technology, Wuhan 430074, China}

\begin{abstract}
Superconducting topological metals (SCTMs) have recently
emerged as a promising platform of topological superconductivity (TSC)
and Majorana zero modes(MZMs) for quantum computation.
Despite their importance in both fundamental research and applications, SCTMs are very rare in nature.
In addition, some superconductors with topological electronic structures have been reported recently,
but a feasible program to determine their TSC properties is still lacking.
Here, we propose a new strategy to design SCTMs by intercalating the superconducting units into the topological insulators.
A program that characterizes the superconducting BdG Chern number of 2D BdG Hamiltonian from ab initio calculations is also developed.
Following this strategy, PdBi$_2$Te$_5$ and PdBi$_2$Te$_4$ are found to be experimentally synthesizable and ideal SCTMs.
Chiral TSC could be realized in such SCTMs by incorporating topological surface states with Zeeman effect,
which can be realized by an external magnetic field or in proximity to ferromagnetic (FM) insulator.
Our strategy provides a new method for identifying the SCTMs and TSC candidates,
and the program makes it possible to design and modulate the TSC candidates from ab initio calculations.

\end{abstract}

\maketitle

\section*{main}
As one of the most important systems in both fundamental physics and topological quantum computation,
topological superconductors (TSCs) have attracted increasing interest for
their ability to support Majorana fermions and anyons with non-Abelian statistics~\cite{Nayak2008, Sarma2015, Sato2017, Lutchyn2018,
Liu2011, Zhang2013, Yang2014, Wang2016, Wang2018b, Hao2019,
Zhang2019b, Zhang2019c, Zhang2021, Wu2020, Giwa2021, Nayak2021, Margalit2022}.
Currently, the search for TSCs candidates has been focused on two experimental schemes.
One is the architectures by the combination of conventional superconductors with
topological insulators (TIs)~\cite{Fu2008,Xu2015,Sun2016}
or 1D nanowires~\cite{Mourik2012, Stevan2014},
but this approach brings high requirements for sample fabrication and interface engineering.
The other route is to achieve TSCs in superconducting topological metals (SCTMs)
that host both topological electronic structures at the Fermi level and superconductivity in one compound~\cite{Hosur2011, Xu2016,
Zhang2018, Wang2018, Liu2018, Liu2020, Kong2021, Li2022, Yuan2019, Fang2019, Li2021, Lv2017, Guan2019, Li2019},
in which the topological surface states are gapped by the ``self-proximity effect" of bulk superconductivity,
thus avoiding the complications of interface engineering.
This approach has successfully predicted the SCTM FeTe$_{0.55}$Se$_{0.45}$~\cite{Xu2016, Zhang2018, Wang2018}
and similar compounds of iron-based superconductors~\cite{Liu2018, Liu2020, Kong2021, Li2022},
owing to the favorable SC gap and non-trivial band topology.
Beside the MZMs, 1D helical/chiral Majorana states have also been reported in domain walls of FeTe$_{0.55}$Se$_{0.45}$~\cite{Wang2020}
and the magnetism-superconductor heterostructures~\cite{Qi2010, Wang2015, He2019,
Zhang2021a, Zhang2021b, Menard2017, Alexandra2019, Kezilebieke2020}.
It is also proposed that the propagating chiral Majorana states
can be applied to realize non-Abelian quantum gate operations,
which could be $10^3$ faster than the currently existing quantum computation schemes~\cite{Biao2018}.

Encouraged by the success of Fe(Se,Te)~\cite{Xu2016, Zhang2018, Wang2018},
many topological materials that host both superconductivity and topological electronic structures are proposed
~\cite{Huang2021,Chen2020,Ortiz2020,Leng2017,Das2018,Kudo2016}.
However, very rare experimental progress of TSC has been made in such SCTMs.
This is because, on the one hand, all of them are not the ideal SCTMs,
whose band structures are too complicated,
the topological surface states are usually buried in the bulk states and difficult to form the pairing required by TSC.
On the other hand,
lacking a direct characterization of the TSC properties from ab initio calculations
also hinders the effective experimental search in such materials.
Therefore, a general program that could calculate the TSC invariant
from first-principles calculations is highly desirable.

In this work, we develop a program to characterize the superconducting topological invariant of 2D system from ab initio calculations.
Besides, we also propose a new strategy to design ideal SCTMs by intercalating superconducting units into topological insulators.
Following this strategy,
PdBi$_2$Te$_5$ and PdBi$_2$Te$_4$ are found to be ideal SCTMs
that host topological surface states at the Fermi level and superconductivity at $0.57$~K and $3.11$~K respectively.
By performing the superconducting energy spectrum and topological invariant calculations,
we identify that chiral TSC could be realized in the slab of such SCTMs
by introducing considerable Zeeman splitting on the topological surface states,
which can be realized by an external magnetic field or in proximity to FM insulators.
Our strategy provides a new framework to enrich SCTMs and TSC candidates,
and the program makes it possible to design and modulate the TSC system from ab initio calculations,
which can also be extended to study the TSC properties in other system,
such as magnetic TI/SC heterostructure, SC/FM heterostructure and SC/TI/SC heterostructure.

\begin{figure}[t]
  \includegraphics[width=0.5\columnwidth]{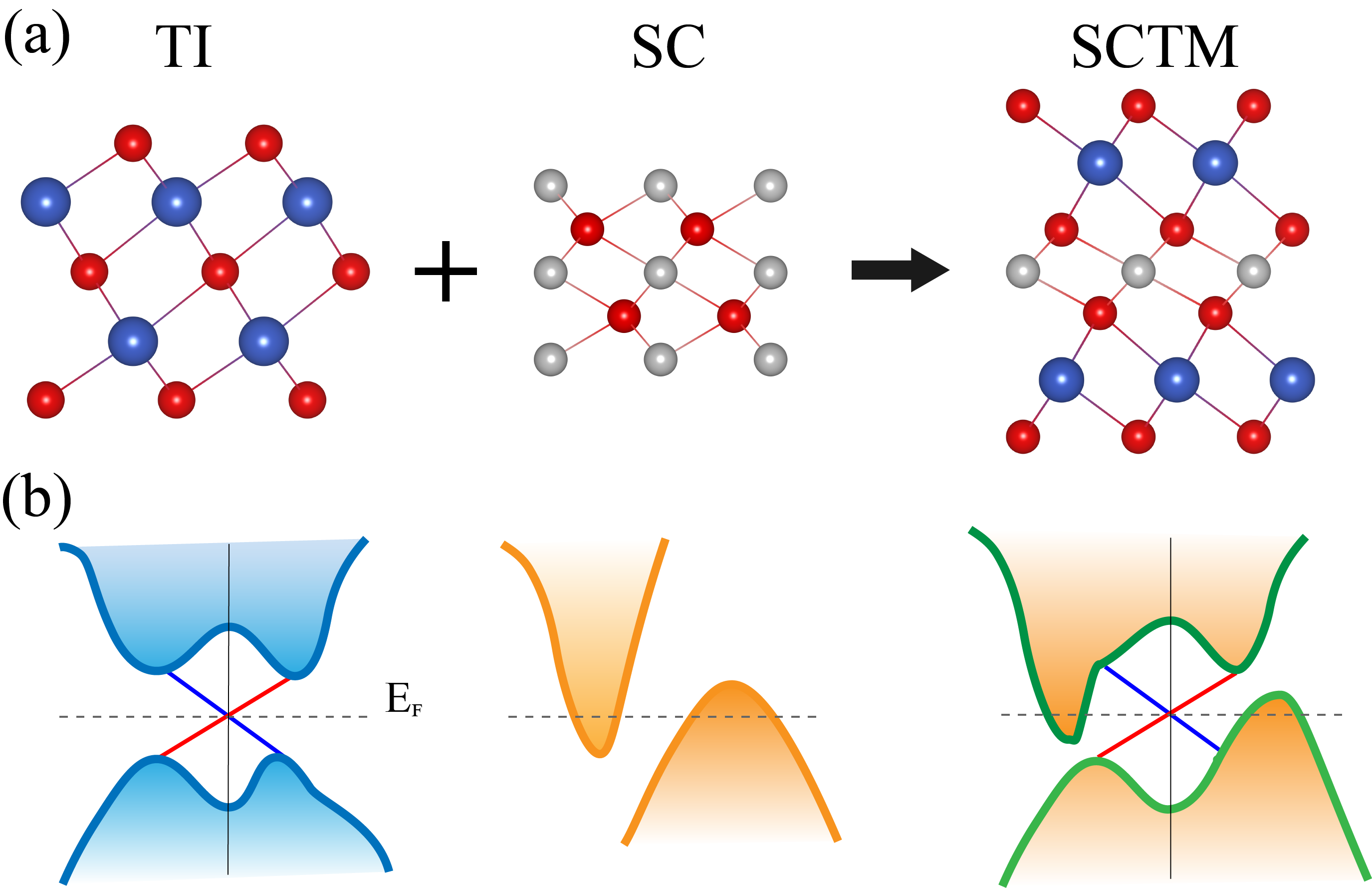}
  \caption{The strategy to design idea SCTMs by intercalating the SC units into the TI.}
  \label{fig:1}
\end{figure}

Inspired by the construction of magnetic TI MnBi$_2$Te$_4$~\cite{Lee2013, Zhang2019},
we propose that the SCTMs can be designed by intercalating the SC units into the TIs,
as illustrated by the schematic of Fig.~\ref{fig:1}(a).
As an ideal SCTM, the target crystal should be relatively stable in both energy and structure.
More importantly, it must inherit the topological electronic structures of the parent TI near the Fermi level,
and also the superconductivity of the parent SC as shown in Fig.~\ref{fig:1}(b).
However, the combination of topological electronic structures and SC does not result in TSC eventually.
The realization of TSC generally requires a delicate modulation of many parameters,
such as SC pairing, Zeeman splitting and chemical potential, et al
~\cite{Fu2008,Xu2015,Sun2016, Mourik2012, Stevan2014, Hosur2011,Xu2016,
Zhang2018, Wang2018, Menard2017, Alexandra2019, Kezilebieke2020, Qi2010, Wang2015, He2019,
Zhang2021a, Zhang2021b}.
Thus, the ability to characterize the TSC invariant and determine the required parameters in real materials
from the ab initio calculations is not only of theoretical significance, but also highly desirable in experiment.

\begin{figure}[t]
  \includegraphics[width=0.6\columnwidth]{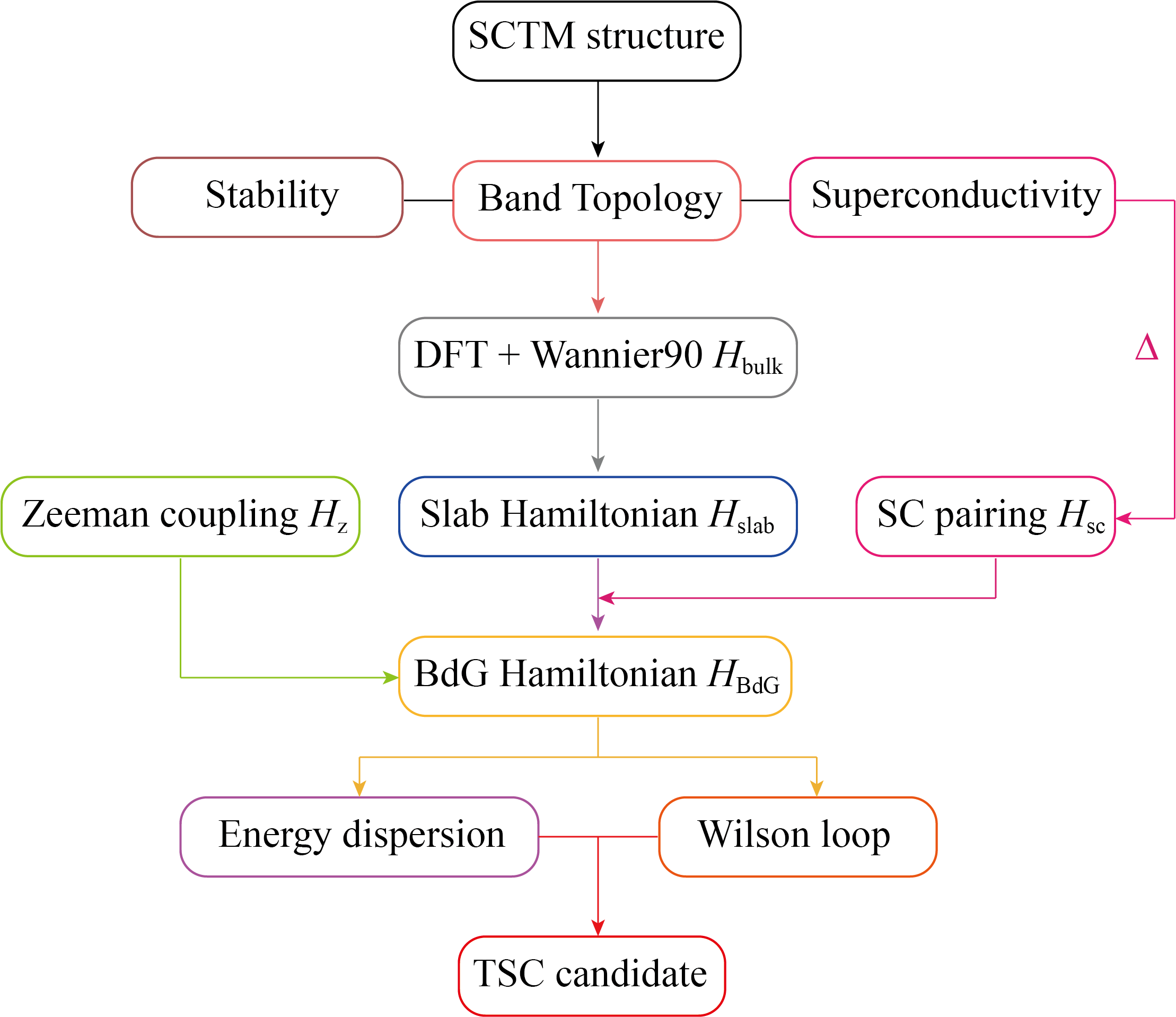}
  \caption{The generic flow chart to characterize the TSC properties from ab initio calculations.}
  \label{fig:2}
\end{figure}

Here we develop a program to simulate the superconducting properties and characterize its
topological invariant in 2D slab system from ab initio calculations,
in which the necessary ingredients to realize chiral TSC based on SCTM are included,
such as bulk band structures, SC pairing, Zeeman splitting, Rashba spin-orbit coupling and chemical potential.
The workflow of this program is shown in Fig.~\ref{fig:2}.
First, one should calculate the electronic structures of SCTM materials,
and construct the localized Wannier functions that capture all electronic features from the first-principles calculations,
referred as $\hat{H}_{\mathrm{bulk}}$.
The next step is to construct the slab Hamiltonian $\hat{H}_{\mathrm{slab}}$
with open boundary condition along a certain direction~\cite{Wu2018}.
In general, the spin-orbit coupling (SOC) and surface effect can be included automatically in $\hat{H}_{\mathrm{slab}}$
through the first-principles calculations with SOC.
So that the topological properties, such as the surface states and spin-texture, can be directly studied by using $\hat{H}_{\mathrm{slab}}$.
On the other hand, one can also construct a slab Hamiltonian $\hat{H}^{\mathrm{nsoc}}_{\mathrm{slab}}$
that excluded SOC from the non-SOC first-principles calculations,
and add $\hat{H}_{\mathrm{SOC}}$ and $\hat{H}_{\mathrm{surf}}$ manually
to simulate the variable SOC and surface effect in the topological electronic states and TSC.
In this work, we will adopt the former type of $\hat{H}_{\mathrm{slab}}$,
in which only the intrinsic SOC of the real material is included.
With adopting particle-hole transformation,
the $\hat{H}_{\mathrm{slab}}$ can be extended to BdG Hamiltonian $\hat{H}^{\mathrm{BdG}}_{\mathrm{slab}}$
by adding SC pairing $\hat{H}_{sc}$ and Zeeman splitting $\hat{H}_{\mathrm{z}}$.
In the Nambu basis $\Phi_\mathbf{k}=(c_{\mathbf{k},j,\alpha, \uparrow}, c_{\mathbf{k},j,\alpha,\downarrow},
c^{\dag}_{-\mathbf{k},j,\alpha,\uparrow}, c^{\dag}_{-\mathbf{k},j,\alpha,\downarrow})$,
where the $c_{j,\alpha,\sigma}$ is the fermion operator denotes an electron at $j$ layer
with orbital $\alpha$ and  spin $\sigma (\uparrow,\downarrow)$,
the BdG Hamiltonian is formulated as:
\begin{equation}\label{eq: PdBiTe BdG}
    H^{\mathrm{BdG}}_{\mathrm{slab}}(\mathbf{k})=
    \begin{pmatrix}
     H_{\mathrm{slab}}(\mathbf{k})-\mu & \Delta(\mathbf{k}) \\
     \Delta^{\dag}(\mathbf{k})  & -H_{\mathrm{slab}}^{*}(-\mathbf{k}) + \mu
    \end{pmatrix}
    + M_{z}\tau_z.
\end{equation}

In Eq.~\ref{eq: PdBiTe BdG},
$\mu$ is the chemical potential, which can be used to simulate the carriers doping.
$\Delta(\mathbf{k})$ denotes the SC pairing matrices, which could be both singlet and triplet pairing form.
For the conventional $s$-wave SC,
$\Delta(\mathbf{k})$ is expressed as:
\begin{equation}\label{eq: PdBi2Te4 pairing slab}
  \Delta(\mathbf{k}) = \Delta_s \times I_{slab}\otimes(i \sigma_y \otimes I_{orb}),
\end{equation}
where $\Delta_s$ is the magnitude of intrinsic bulk $s$-wave pairing,
$\sigma_y$ is the Pauli matrix in spin space,
$I_{slab}$ ($I_{orb}$) is an $\mathcal{N}_{slab}\times\mathcal{N}_{slab}$ ($\mathcal{N}_{orb}\times\mathcal{N}_{orb}$)
identity matrix that represents the number of slab layers (Wannier orbitals).
$\tau_z$ is the Pauli matrix in particle-hole space,
$M_z$ is the Zeeman splitting energy,
and $H_z = M_z \tau_z$ is used to simulate the influence of the external magnetic field
or the proximity effect of the FM insulator.
Thus, $H_z$ can be chosen to be applied for the whole slab or just few surface layers,
depending on the slab thickness, strength of magnetic field, the type of the SC et al.
In principle, chiral TSC can be achieved by modulating the SC pairing,
Zeeman splitting and chemical potential~\cite{Menard2017, Alexandra2019, Kezilebieke2020, Qi2010, Wang2015, He2019, Zhang2021a, Zhang2021b},
which can be further revealed by calculating the superconducting energy spectrum and the superconducting topological invariant.

In the gaped 2D superconducting system,
the  topological superconductors are classified by BdG Chern number
in the absence of time-reversal symmetry~\cite{Sato2017}.
Such superconducting topological invariants can be characterized
by the evolution of Wilson loop~\cite{Yu2011, Soluyanov2011, Gresch2017}.
For the occupied quasiparticle states $|u^{\textrm{BdG}}_{n,k_1,k_2} \rangle$,
where $k_1$ and $k_2$ are momenta along two primitive vectors of the
Brillouin zone (BZ),
the Berry phase of the Wilson loop along $k_2$ at a fixed $k_1$ can be expressed as:
\begin{equation}\label{eq:BdG Wilson loop}
  \mathcal{W}(k_1)=-\mathrm{Im} \ln \prod_{i} \det M^{(i)}_{k_1},
\end{equation}
with the overlap matrix $M^{(i)}_{k_1,mn}=\langle u^{\textrm{BdG}}_{m,k_1,k^{(i)}_2} |u^{\textrm{BdG}}_{n,k_1, k^{(i+1)}_2} \rangle$,
where $k^{(i)}_2$ is the $i$-th discretized momenta along $k_2$ direction.
The winding number of $\mathcal{W}(k_1)$ with respect to $k_1$
is equal to the superconducting BdG Chern number $C_{\textrm{BdG}}$.

Next, we take TI Bi$_2$Te$_3$~\cite{Zhang2009,Chen2009}, SC PdTe~\cite{Matthias1953,Karki2012} and SC PdTe$_2$~\cite{Leng2017,Das2018,Kudo2016}
as parent compounds to demonstrate that our SCTMs strategy is feasible.
Experimentally, Bi$_2$Te$_3$ (space group $R\bar{\mathrm{3}}m$, $a=4.35$~\AA, $c=30.36$~\AA),
PdTe (space group space group $P\rm{6_3}/mmc$, a = 4.152~\AA, c = 5.671~\AA, T$_c$ = 2.3~K)
and PdTe$_2$ (space group $P\bar{\mathrm{3}}m$1, $a=4.03$~\AA, $c=5.12$~\AA, T$_c=1.64$~K)
all adopt the triangle lattice
and have very similar in-plane lattice constants,
which makes it much easier to integrate them together to form a new compound.
According to our calculations,
the stable unit of PdBi$_2$Te$_5$ and PdBi$_2$Te$_4$
adopt octuple-layer (OL) structure and septuple-layer (SL) structure respectively,
as shown in Fig.~\ref{fig:3}(a)(also Fig.~S1) and Fig.~S2 of Supplementary Material(SM)~\cite{SM}.
They both favor the ABC stacking along c-direction,
and form the rhombohedral unit cell as shown in Fig.~\ref{fig:3}(a),
which is 73~meV/f.u. (73~meV/f.u. for PdBi$_2$Te$_4$) and 46~meV/f.u. (12~meV/f.u. PdBi$_2$Te$_4$)
lower than the AA and AB stacking structures.
The detailed crystal parameters and total energy of different stacking PbBi$_2$Te$_5$ and PbBi$_2$Te$_4$
are tabulated in the Table.~S1 and Table.~S2, respectively~\cite{SM}.

The formation energy of PdBi$_2$Te$_5$ and PdBi$_2$Te$_4$ are
calculated to study their thermodynamic stability
by using
$E_{f}^{\mathrm{Pd_mBi_nTe_l}}
= E^{\mathrm{Pd_mBi_nTe_l}} - mE^{\mathrm{Pd}} - nE^{\mathrm{Bi}} - lE^{\mathrm{Te}}$,
with $E^{i}$($i$=$\mathrm{Pd_mBi_nTe_l}, \mathrm{Pd}, \mathrm{Bi}~ \mathrm{and}~\mathrm{Te}$) means the calculated total energy per formula in the ground state.
The calculated $E^{PdBi_2Te_5}_f$ and $E^{PdBi_2Te_4}_f$
are $-3.184$~eV/f.u. and $-2.476$~eV/f.u.,
which means that 3.184~eV and 2.476~eV can be released during their synthesis processes
from the constituent elements.
To further manifest their thermodynamic stability,
we construct the convex hull diagram in Fig.~\ref{fig:3}(b)
with all of the synthesized Pd-Bi-Te compounds,
whose crystal parameters and the calculated formation energy have been tabulated in Table.~S3 and Table.~S4, respectively~\cite{SM}.
Fig.~\ref{fig:3}(b) shows that
PdBi$_2$Te$_5$ and PdBi$_2$Te$_4$ are 13~meV/atom and 61~meV/atom above the convex hull respectively.
Moreover, considering that metastable PdBi$_2$Te$_3$,
52~meV and 3~meV higher than PdBi$_2$Te$_5$ and PdBi$_2$Te$_4$ as shown in Fig.~\ref{fig:3}(b),
has been synthesized in experiments~\cite{Sharma2020,Wang2021},
we thus conclude that PdBi$_2$Te$_5$ and PdBi$_2$Te$_4$ could be synthesized in experiments.
For PdBi$_2$Te$_5$,
we propose a synthetic route through the growth of Bi$_2$Te$_3$ and PdTe$_2$ layer by layer.
Our calculated results reveal that
bulk PdBi$_2$Te$_5$ is 59~meV/f.u. lower than the total energy of free standing Bi$_2$Te$_3$ and PdTe$_2$ layers,
which strongly suggest that PdTe$_2$ layer tends to deposit on Bi$_2$Te$_3$ to form new PdBi$_2$Te$_5$ crystal.
To investigate their dynamical stability,
we calculate the phonon dispersion of PdBi$_2$Te$_5$ and PdBi$_2$Te$_4$,
and plot them in Fig.~\ref{fig:3}(c) and Fig.~S3(a)~\cite{SM}.
There are 24 (21) phonon modes with fully real positive frequencies for PdBi$_2$Te$_5$ (PdBi$_2$Te$_4$),
which indicates that the rhombohedral unit cells are dynamically stable.
Based on these results, we conclude that PdBi$_2$Te$_5$ and PdBi$_2$Te$_4$
are relatively thermodynamically and dynamically stability
in the rhombohedral structure,
and further experimental investigation is called for.

Then we study the electronic structures and topological properties of PdBi$_2$Te$_5$ and PdBi$_2$Te$_4$.
Since PdBi$_2$Te$_5$ and PbBi$_2$Te$_4$
exhibit similar electronic structures and non-trivial band topology,
we only show the detailed density of states (DOS), band structures, and topological surface states of PdBi$_2$Te$_5$ as an example in the main text,
one can check the results of PdBi$_2$Te$_4$ in Section III and Figs.~S3 of the SM~\cite{SM}.
In Fig.~\ref{fig:3}(d), we plot the total and projected DOS of PdBi$_2$Te$_5$,
which gives rise to DOS(0 eV) = 1.91~states/eV at Fermi level, indicating its metallic nature and the possibility of superconductivity.
The projected DOS demonstrates that the states between $-1$ eV and $1$ eV
are dominated by the $p$-orbitals of Te hybridized with $d$-orbitals from Pd and $p$-orbitals from Bi.
The hybridization is also manifested by the projected band structures shown in Fig.~\ref{fig:3}(e),
which shows that two bands with $p$-orbital components from Te or Bi cross the Fermi level and form several Fermi surfaces.
Further detailed orbital components analysis demonstrates that a continuous band gap (yellow region in Fig.~\ref{fig:3}(e))
and band inversion exists between the nominal valence band and conduction band around the Fermi level,
which implies that PdBi$_2$Te$_5$ inherits the topological electronic nature of Bi$_2$Te$_3$ successfully.
The nontrivial band topology can be confirmed by calculating the $Z_2$ topological invariant of time-reversal invariant insulators~\cite{Fu2007}.
Given that rhombohedral PdBi$_2$Te$_5$ possesses inversion symmetry and a continuous band gap,
the $Z_2$ topological invariant
$\nu_{\text{TI}}=(1-P)/2$ is determined by the product $P$ of the parity of
the wave function at the TRIM points~\cite{Fu2007}.
Our calculated results give $Z_2$ index $\nu_{\text{TI}}=1$,
confirming PdBi$_2$Te$_5$ is a $Z_2$ topological metal.
To visualize the bulk–boundary correspondence,
we calculate and plot the topological surface states on the (001) surface in Fig.~\ref{fig:3}(f).
The surface states are similar to that of Bi$_2$Te$_3$~\cite{Zhang2009,Chen2009},
the Dirac cone at the $\Gamma$ point manifest approximately $-6.3$~meV below the Fermi level (the dashed line in Fig.~\ref{fig:3}(f)).

To investigate  the superconducting property of PdBi$_2$Te$_5$,
we perform the electron-phonon calculations based on density functional perturbation theory~\cite{Baroni2001}.
The calculated electron-phonon coupling constant $\lambda=0.43$ and logarithmic average phonon frequency $\omega_{\log}=97~cm^{-1}$,
as tabulated in Table.~S5~\cite{SM}.
Furthermore, the superconducting transition temperature ($T_c$) is estimated by using the reduced Allen-Dynes formula~\cite{McMillan1968,Allen1975}:
\begin{equation}
 T_{c}=\frac{\omega_{\log}}{1.20} \exp{\left[-\frac{1.04(1+\lambda)}{\lambda-\mu^*(1+0.62\lambda)}\right]},
\end{equation}
where $\mu^*$ is the effective Coulomb potential.
By adopting a typical $\mu^*$ = 0.1,
the $T_c$ of PdBi$_2$Te$_5$ is estimated as 0.57~K.
As comparison, the calculated $\lambda$ and $\omega_{\log}$ in PdTe$_2$ is 0.52 and 112 $cm^{-1}$, respectively.
Accordingly, the estimated $T_c$ in PdTe$_2$ is 1.59~K,
which agrees well with the experimental $T_c$ of 1.64~K~\cite{Leng2017,Das2018,Kudo2016}.
These results clearly demonstrate that the SC in PdTe$_2$ is well inherited into the PdBi$_2$Te$_5$.

\begin{figure}[t]
  \includegraphics[width=0.6\columnwidth]{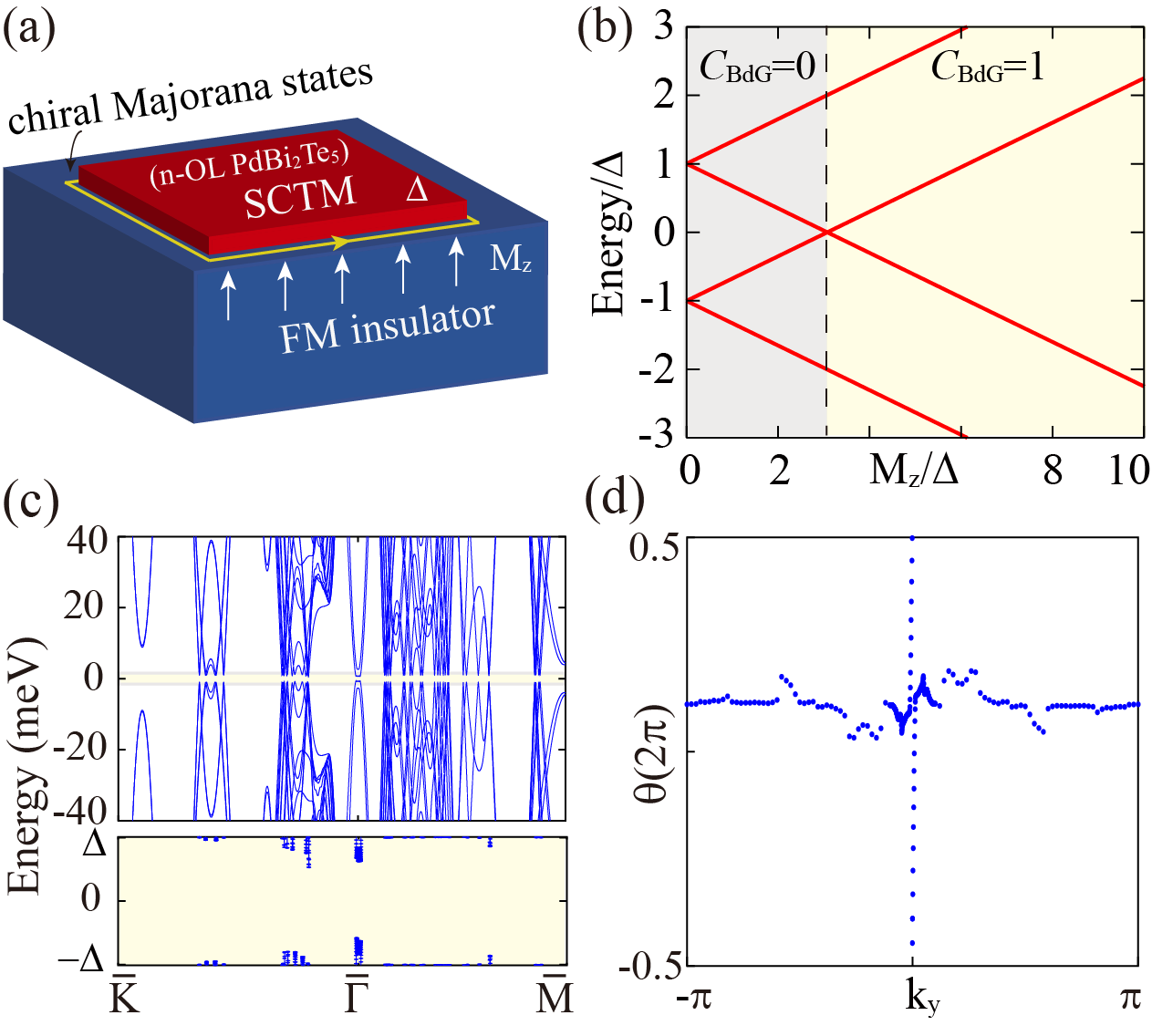}
  \caption{(a) The schematic to realize chiral TSC in PdBi$_2$T$_5$ slab.
  (b) The low energy spectrum at the $\Gamma$ point as the function of Zeeman splitting energy $M_z$.
  (c) The superconducting spectrum along high symmetry paths with $M_z=5$~meV and $\Delta=1$~meV,
  the zoom-in image shows the full gap in the whole BZ.
  (d) The Wilson loop spectrum for all occupied states of 4(c), which manifest the superconducting BdG Chern number $C_{\textrm{BdG}}=1$ clearly.}
  \label{fig:4}
\end{figure}

We now study the TSC property of the PdBi$_2$Te$_5$ slab
by introducing the SC pairing and Zeeman splitting into the topological surface states.
Usually, the Zeeman splitting is applied by external magnetic field or in proximity to a FM insulator,
as illustrated in Fig.~\ref{fig:4}(a).
As a concrete example, we use a 2D slab consisting of 10-OL PdBi$_2$Te$_5$,
which is thick enough to avoid the hybridization between top layer and bottom layer (Fig.~\ref{fig:3}(f)).
Since PdBi$_2$Te$_5$ is an intrinsic SC,
the estimated $s$-wave superconducting gap $\Delta_s=1.0$~meV
is introduced globally for all 10-OLs.
The out-of-plane Zeeman splitting is applied only in the bottom layer consisting of one Bi$_2$Te$_3$ and one PdTe$_2$,
by assuming PdBi$_2$Te$_5$ is the conventional SC from the parent type-I SC PdTe$_2$~\cite{Leng2017}.
The chemical potential $\mu$ is set at the energy
of surface Dirac cone at the $\Gamma$ point
(about $-6.3$ meV below the Fermi level).
In Fig.~\ref{fig:4}(b), we show the low energy spectrum of $H_{\mathrm{BdG}}$ at $\Gamma$
point as a function of Zeeman splitting energy $M_z$,
which manifest that the superconducting spectrum is fully gaped with an energy gap of $\Delta$ at $M_z=0$.
As $M_z$ increases, the superconducting gap at the $\Gamma$ point closes and reopens.
This behavior indicates that a topological phase transition happens at critical point $M_z/\Delta=3.1$,
and this 2D slab enters chiral TSC phase characterized
by a nonzero BdG Chern number and chiral Majorana edge states according to previous model simulations~\cite{Qi2010, Wang2015, He2019}.

To firmly verify its topological property and visualize the low energy physics in the TSC phase,
we calculate the superconducting energy spectrum at $M_z$=5 meV and $\Delta$=1 meV in Fig.~\ref{fig:4}(c).
The corresponding Wilson loop evolutions for the occupied states are ploted in Fig.~\ref{fig:4}(d).
The zoom-in image of Fig.~\ref{fig:4}(c) reveals that a full superconducting gap is opened in the whole BZ,
indicating that the system is a well defined chiral TSC.
The Wilson loop evolution exhibits a nontrivial chiral winding number 1,
which directly confirms the superconducting BdG Chern number $C_{\textrm{BdG}}=1$.
Given that the experimental accessible magnetization energy usually reaches a few tens of meV,
our results provide a feasible guideline for discovery the chiral TSC phase in PdBi$_2$Te$_5$.

Finally, we would like to point out that the chiral TSC phase could also be realized in PdBi$_2$Te$_4$
as shown in Fig.~S4~\cite{SM}, which exhibits a similar superconducting spectrum gap closing behavior with respect to $M_z/\Delta$ as in PdBi$_2$Te$_5$.
In addition, we emphasize that our material design strategy can also be applied to search for other SCTM candidates.
For example,
our calculated results demonstrate that AuBi$_2$Te$_5$ formed by SC AuTe$_2$ interacting into Bi$_2$Te$_3$ is also an ideal SCTM,
whose detailed crystal structures, dynamic stability, electronic structures, and topological surface states are discussed in Section V and Fig.~S5 of SM~\cite{SM}.
Therefore, we expect that SCTM AuBi$_2$Te$_5$ could also be a TSC candidate.
Last, we would like to point out that the program
can be extended to study many 2D topological superconducting heterostructure systems,
such as magnetic TI/SC heterostructure,
SC/FM heterostructure and SC/TI/SC heterostructure.
This will make it possible to determine the accurate parameters of the TSC phase and simulate their TSC property in such systems from first-principles calculations.
We expect our program to be also useful for optimizing the experimental setup,
stimulating the field of TSC study.

\section*{Acknowledgments}
This work is supported by the National Key Research and Development Program of China (2018YFA0307000),
and the National Natural Science Foundation of China (12274154, 11874022).
B.L. is supported by the Alfred P. Sloan Foundation, the National Science Foundation through Princeton University’s Materials Research Science and Engineering Center DMR-2011750, and the National Science Foundation under award DMR-2141966.

%\section*{Author contributions}
%G. X. conceived and designed the project. A. L. Y. L. and Y. Q. performed the DFT calculations.
%G. X., B. L. and A. L. did the theoretical analysis.
%All authors contributed to the manuscript writing.
%
%\section*{Competing Interests}
%The authors declare no competing financial interests.
%
%
%\section*{Correspondence}
%Correspondence and requests for materials should be addressed to G. Xu~(email: gangxu@hust.edu.cn).

\section*{Method}
The first-principles calculations based on density functional theory
are performed by the Vienna ab initio simulation package~\cite{Kresse1993,Kresse1996}
with treating Perdew–Burke–Ernzerhof type of generalized gradient approximation as the exchange-correlation potential~\cite{Perdew1996}.
The cutoff energy for wave function expansion is set as 450~eV,
$k$-points grid 13$\times$13$\times$13 is used for sampling the first BZ.
All crystal structures are fully optimized until the force on each atom is less than 0.01 eV/\AA,
and the SOC is included self-consistently.
The electron-phonon coupling calculations with
van der Waals correction~\cite{Klime2011}
are carried out
in Quantum Espresso~\cite{Giannozzi2009} based on the perturbation theory.
A Hermite-Gaussian smearing of 0.0025 Ryd is used for the electronic integration.
The 8$\times$8$\times$8 $k$-mesh is used for the electron-phonon coupling strength $\lambda$ calculations,
and the dynamical matrices are calculated on a 4$\times$4$\times$4 phonon-momentum grid.
Besides, a 2$\times$2$\times$2 supercell is built to calculate the phonon dispersion by using PHONOPY~\cite{Togo2015}.
For the surface calculation, the Wannier functions of Pd-$d$, Bi-$p$ and Te-$p$ orbitals are constructed by using WANNIER90~\cite{Mostofi2008}.
A slab consisting of 10-OL PdBi$_2$Te$_5$ layers with a bottom surface terminated as the Bi$_2$Te$_3$ layer is implemented in WannierTools~\cite{Wu2018},
which is further used to calculate the electronic surface states,
the superconducting spectrum, and the superconducting BdG Chern number.

\newpage

\begin{figure}[t]
  \includegraphics[width=1.0\columnwidth]{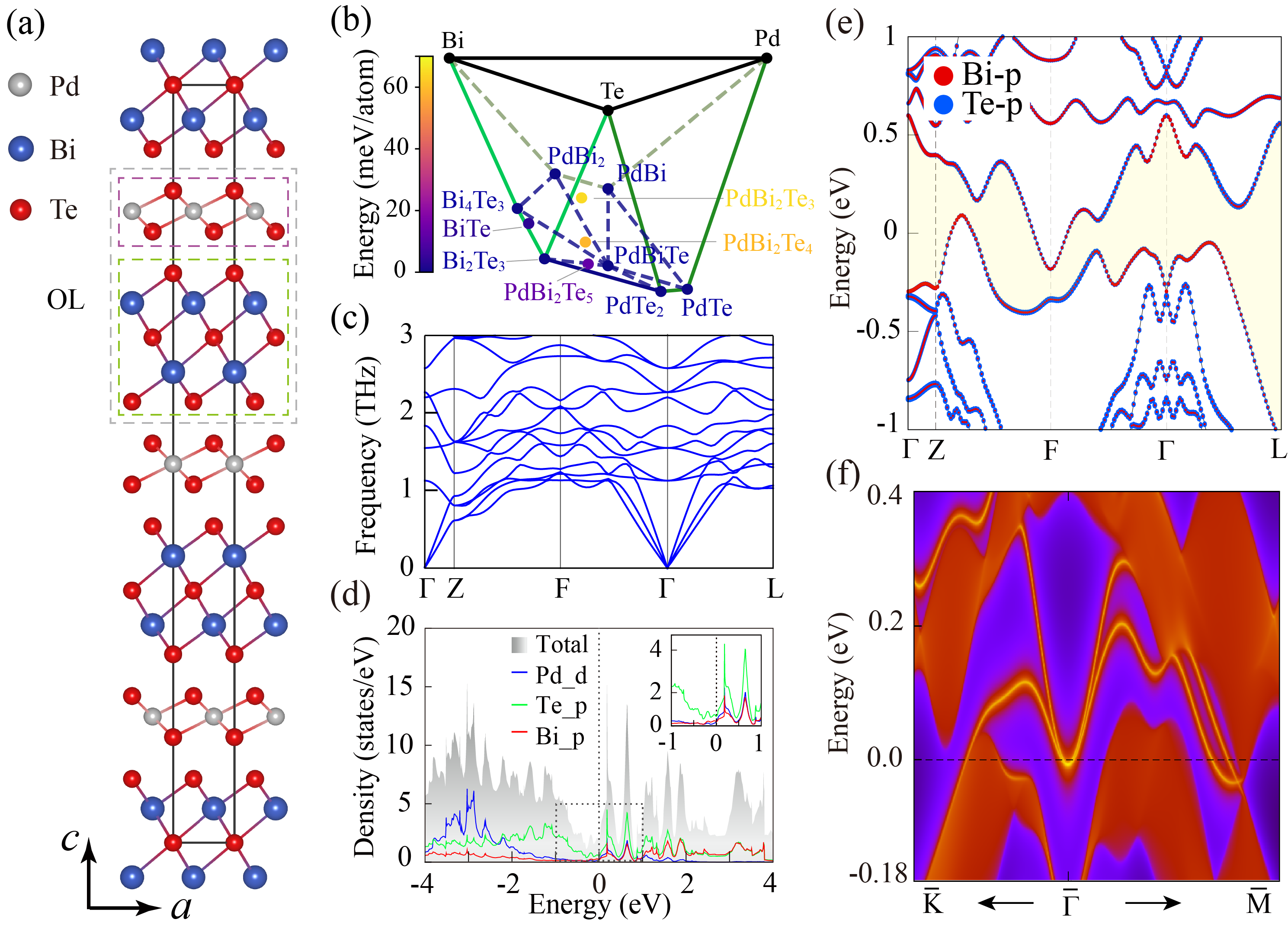}
  \caption{(a) The side view of the crystal structures of PdBi$_2$Te$_5$,
  in which the octuple-layer (OL) unit of PdBi$_2$Te$_5$ formed by the Bi$_2$Te$_3$ quintuple-layer and PdTe$_2$ triple-layer is marked by a grey dashed rectangle.
  (b) Convex hull diagram for Pd-Bi-Te system, the energy above convex hull is displayed by color-bar.
  (c) The phonon dispersion of PdBi$_2$Te$_5$.
  (d) The total DOS and projected DOS of the Pd, Te, Bi atoms in PdBi$_2$Te$_5$,
  the zoom-in image shows the projected DOS near the Fermi level.
  (e) The orbital-projected band structures of PdBi$_2$Te$_5$, where a continuous band gap around the Fermi level is marked by the yellow shade.
  (f) The topological surface states on (001) surface of PdBi$_2$Te$_5$.}
  \label{fig:3}
\end{figure}

\begin{thebibliography}{10}
\expandafter\ifx\csname url\endcsname\relax
  \def\url#1{\texttt{#1}}\fi
\expandafter\ifx\csname urlprefix\endcsname\relax\def\urlprefix{URL }\fi
\providecommand{\bibinfo}[2]{#2}
\providecommand{\eprint}[2][]{\url{#2}}
\bibitem{Nayak2008}
\bibinfo{author}{Nayak, C.}, \bibinfo{author}{Simon, S.~H.},
  \bibinfo{author}{Stern, A.}, \bibinfo{author}{Freedman, M.} \&
  \bibinfo{author}{Das~Sarma, S.}
\newblock \bibinfo{title}{Non-abelian anyons and topological quantum
  computation}.
\newblock \emph{\bibinfo{journal}{Rev. Mod. Phys.}}
  \textbf{\bibinfo{volume}{80}}, \bibinfo{pages}{1083--1159}
  (\bibinfo{year}{2008}).

\bibitem{Sarma2015}
\bibinfo{author}{Sarma, S.~D.}, \bibinfo{author}{Freedman, M.} \&
  \bibinfo{author}{Nayak, C.}
\newblock \bibinfo{title}{Majorana zero modes and topological quantum
  computation}.
\newblock \emph{\bibinfo{journal}{npj Quantum Information}}
  \textbf{\bibinfo{volume}{1}}, \bibinfo{pages}{1--13} (\bibinfo{year}{2015}).

\bibitem{Sato2017}
\bibinfo{author}{Sato, M.} \& \bibinfo{author}{Ando, Y.}
\newblock \bibinfo{title}{Topological superconductors: a review}.
\newblock \emph{\bibinfo{journal}{Reports on Progress in Physics}}
  \textbf{\bibinfo{volume}{80}}, \bibinfo{pages}{076501}
  (\bibinfo{year}{2017}).

\bibitem{Lutchyn2018}
\bibinfo{author}{Lutchyn, R.~M.} \emph{et~al.}
\newblock \bibinfo{title}{Majorana zero modes in superconductor--semiconductor
  heterostructures}.
\newblock \emph{\bibinfo{journal}{Nat. Rev. Mater.}}
  \textbf{\bibinfo{volume}{3}}, \bibinfo{pages}{52--68} (\bibinfo{year}{2018}).

\bibitem{Liu2011}
\bibinfo{author}{Liu, C.-X.} \& \bibinfo{author}{Trauzettel, B.}
\newblock \bibinfo{title}{Helical Dirac-Majorana interferometer in a
  superconductor/topological insulator sandwich structure}.
\newblock \emph{\bibinfo{journal}{Phys. Rev. B}} \textbf{\bibinfo{volume}{83}},
  \bibinfo{pages}{220510} (\bibinfo{year}{2011}).

\bibitem{Zhang2013}
\bibinfo{author}{Zhang, F.}, \bibinfo{author}{Kane, C.~L.} \&
  \bibinfo{author}{Mele, E.~J.}
\newblock \bibinfo{title}{Time-reversal-invariant topological superconductivity
  and Majorana Kramers pairs}.
\newblock \emph{\bibinfo{journal}{Phys. Rev. Lett.}}
  \textbf{\bibinfo{volume}{111}}, \bibinfo{pages}{056402}
  (\bibinfo{year}{2013}).

\bibitem{Yang2014}
\bibinfo{author}{Yang, S.~A.}, \bibinfo{author}{Pan, H.} \&
  \bibinfo{author}{Zhang, F.}
\newblock \bibinfo{title}{Dirac and Weyl superconductors in three dimensions}.
\newblock \emph{\bibinfo{journal}{Phys. Rev. Lett.}}
  \textbf{\bibinfo{volume}{113}}, \bibinfo{pages}{046401}
  (\bibinfo{year}{2014}).

\bibitem{Wang2016}
\bibinfo{author}{Wang, Q.-Z.} \& \bibinfo{author}{Liu, C.-X.}
\newblock \bibinfo{title}{Topological nonsymmorphic crystalline
  superconductors}.
\newblock \emph{\bibinfo{journal}{Phys. Rev. B}} \textbf{\bibinfo{volume}{93}},
  \bibinfo{pages}{020505} (\bibinfo{year}{2016}).

\bibitem{Wang2018b}
\bibinfo{author}{Wang, Q.}, \bibinfo{author}{Liu, C.-C.}, \bibinfo{author}{Lu,
  Y.-M.} \& \bibinfo{author}{Zhang, F.}
\newblock \bibinfo{title}{High-temperature Majorana corner states}.
\newblock \emph{\bibinfo{journal}{Phys. Rev. Lett.}}
  \textbf{\bibinfo{volume}{121}}, \bibinfo{pages}{186801}
  (\bibinfo{year}{2018}).

\bibitem{Hao2019}
\bibinfo{author}{Hao, N.} \& \bibinfo{author}{Hu, J.}
\newblock \bibinfo{title}{Topological quantum states of matter in iron-based
  superconductors: from concept to material realization}.
\newblock \emph{\bibinfo{journal}{National Science Review}}
  \textbf{\bibinfo{volume}{6}}, \bibinfo{pages}{213--226}
  (\bibinfo{year}{2019}).

\bibitem{Zhang2019b}
\bibinfo{author}{Zhang, R.-X.}, \bibinfo{author}{Cole, W.~S.} \&
  \bibinfo{author}{Das~Sarma, S.}
\newblock \bibinfo{title}{Helical hinge Majorana modes in iron-based
  superconductors}.
\newblock \emph{\bibinfo{journal}{Phys. Rev. Lett.}}
  \textbf{\bibinfo{volume}{122}}, \bibinfo{pages}{187001}
  (\bibinfo{year}{2019}).

\bibitem{Zhang2019c}
\bibinfo{author}{Zhang, R.-X.}, \bibinfo{author}{Cole, W.~S.},
  \bibinfo{author}{Wu, X.} \& \bibinfo{author}{Das~Sarma, S.}
\newblock \bibinfo{title}{Higher-order topology and nodal topological
  superconductivity in Fe(Se,Te) heterostructures}.
\newblock \emph{\bibinfo{journal}{Phys. Rev. Lett.}}
  \textbf{\bibinfo{volume}{123}}, \bibinfo{pages}{167001}
  (\bibinfo{year}{2019}).

\bibitem{Zhang2021}
\bibinfo{author}{Zhang, R.-X.} \& \bibinfo{author}{Das~Sarma, S.}
\newblock \bibinfo{title}{Intrinsic time-reversal-invariant topological
  superconductivity in thin films of iron-based superconductors}.
\newblock \emph{\bibinfo{journal}{Phys. Rev. Lett.}}
  \textbf{\bibinfo{volume}{126}}, \bibinfo{pages}{137001}
  (\bibinfo{year}{2021}).

\bibitem{Wu2020}
\bibinfo{author}{Wu, X.} \emph{et~al.}
\newblock \bibinfo{title}{Boundary-obstructed topological
  high-${\mathit{T}}_{c}$ superconductivity in iron pnictides}.
\newblock \emph{\bibinfo{journal}{Phys. Rev. X}} \textbf{\bibinfo{volume}{10}},
  \bibinfo{pages}{041014} (\bibinfo{year}{2020}).

\bibitem{Giwa2021}
\bibinfo{author}{Giwa, R.} \& \bibinfo{author}{Hosur, P.}
\newblock \bibinfo{title}{Fermi arc criterion for surface Majorana modes in
  superconducting time-reversal symmetric Weyl semimetals}.
\newblock \emph{\bibinfo{journal}{Phys. Rev. Lett.}}
  \textbf{\bibinfo{volume}{127}}, \bibinfo{pages}{187002}
  (\bibinfo{year}{2021}).

\bibitem{Nayak2021}
\bibinfo{author}{Nayak, A.~K.} \emph{et~al.}
\newblock \bibinfo{title}{Evidence of topological boundary modes with
  topological nodal-point superconductivity}.
\newblock \emph{\bibinfo{journal}{Nature Physics}}
  \textbf{\bibinfo{volume}{17}}, \bibinfo{pages}{1413--1419}
  (\bibinfo{year}{2021}).

\bibitem{Margalit2022}
\bibinfo{author}{Margalit, G.}, \bibinfo{author}{Yan, B.},
  \bibinfo{author}{Franz, M.} \& \bibinfo{author}{Oreg, Y.}
\newblock \bibinfo{title}{Chiral Majorana modes via proximity to a twisted
  cuprate bilayer}.
\newblock \emph{\bibinfo{journal}{Phys. Rev. B}}
  \textbf{\bibinfo{volume}{106}}, \bibinfo{pages}{205424}
  (\bibinfo{year}{2022}).

\bibitem{Fu2008}
\bibinfo{author}{Fu, L.} \& \bibinfo{author}{Kane, C.~L.}
\newblock \bibinfo{title}{Superconducting proximity effect and Majorana
  fermions at the surface of a topological insulator}.
\newblock \emph{\bibinfo{journal}{Phys. Rev. Lett.}}
  \textbf{\bibinfo{volume}{100}}, \bibinfo{pages}{096407}
  (\bibinfo{year}{2008}).

\bibitem{Xu2015}
\bibinfo{author}{Xu, J.-P.} \emph{et~al.}
\newblock \bibinfo{title}{Experimental detection of a Majorana mode in the core
  of a magnetic vortex inside a topological insulator-superconductor
  ${\mathrm{Bi}}_{2}{\mathrm{Te}}_{3}/{\mathrm{NbSe}}_{2}$ heterostructure}.
\newblock \emph{\bibinfo{journal}{Phys. Rev. Lett.}}
  \textbf{\bibinfo{volume}{114}}, \bibinfo{pages}{017001}
  (\bibinfo{year}{2015}).

\bibitem{Sun2016}
\bibinfo{author}{Sun, H.-H.} \emph{et~al.}
\newblock \bibinfo{title}{Majorana zero mode detected with spin selective
  andreev reflection in the vortex of a topological superconductor}.
\newblock \emph{\bibinfo{journal}{Phys. Rev. Lett.}}
  \textbf{\bibinfo{volume}{116}}, \bibinfo{pages}{257003}
  (\bibinfo{year}{2016}).

\bibitem{Mourik2012}
\bibinfo{author}{Mourik, V.} \emph{et~al.}
\newblock \bibinfo{title}{Signatures of Majorana fermions in hybrid
  superconductor-semiconductor nanowire devices}.
\newblock \emph{\bibinfo{journal}{Science}} \textbf{\bibinfo{volume}{336}},
  \bibinfo{pages}{1003--1007} (\bibinfo{year}{2012}).

\bibitem{Stevan2014}
\bibinfo{author}{Nadj-Perge, S.} \emph{et~al.}
\newblock \bibinfo{title}{Observation of Majorana fermions in ferromagnetic
  atomic chains on a superconductor}.
\newblock \emph{\bibinfo{journal}{Science}} \textbf{\bibinfo{volume}{346}},
  \bibinfo{pages}{602--607} (\bibinfo{year}{2014}).

\bibitem{Hosur2011}
\bibinfo{author}{Hosur, P.}, \bibinfo{author}{Ghaemi, P.},
  \bibinfo{author}{Mong, R. S.~K.} \& \bibinfo{author}{Vishwanath, A.}
\newblock \bibinfo{title}{Majorana modes at the ends of superconductor vortices
  in doped topological insulators}.
\newblock \emph{\bibinfo{journal}{Phys. Rev. Lett.}}
  \textbf{\bibinfo{volume}{107}}, \bibinfo{pages}{097001}
  (\bibinfo{year}{2011}).

\bibitem{Xu2016}
\bibinfo{author}{Xu, G.}, \bibinfo{author}{Lian, B.}, \bibinfo{author}{Tang,
  P.}, \bibinfo{author}{Qi, X.-L.} \& \bibinfo{author}{Zhang, S.-C.}
\newblock \bibinfo{title}{Topological superconductivity on the surface of
  Fe-based superconductors}.
\newblock \emph{\bibinfo{journal}{Phys. Rev. Lett.}}
  \textbf{\bibinfo{volume}{117}}, \bibinfo{pages}{047001}
  (\bibinfo{year}{2016}).

\bibitem{Zhang2018}
\bibinfo{author}{Zhang, P.} \emph{et~al.}
\newblock \bibinfo{title}{Observation of topological superconductivity on the
  surface of an iron-based superconductor}.
\newblock \emph{\bibinfo{journal}{Science}} \textbf{\bibinfo{volume}{360}},
  \bibinfo{pages}{182--186} (\bibinfo{year}{2018}).

\bibitem{Wang2018}
\bibinfo{author}{Wang, D.} \emph{et~al.}
\newblock \bibinfo{title}{Evidence for Majorana bound states in an iron-based
  superconductor}.
\newblock \emph{\bibinfo{journal}{Science}} \textbf{\bibinfo{volume}{362}},
  \bibinfo{pages}{333--335} (\bibinfo{year}{2018}).

\bibitem{Liu2018}
\bibinfo{author}{Liu, Q.} \emph{et~al.}
\newblock \bibinfo{title}{Robust and clean Majorana zero mode in the vortex
  core of high-temperature superconductor (Li$_{0.84}$Fe$_{0.16}$)OHFeSe}.
\newblock \emph{\bibinfo{journal}{Physical Review X}}
  \textbf{\bibinfo{volume}{8}}, \bibinfo{pages}{041056} (\bibinfo{year}{2018}).

\bibitem{Liu2020}
\bibinfo{author}{Liu, W.} \emph{et~al.}
\newblock \bibinfo{title}{A new majorana platform in an Fe-As bilayer
  superconductor}.
\newblock \emph{\bibinfo{journal}{Nature Communications}}
  \textbf{\bibinfo{volume}{11}}, \bibinfo{pages}{1--7} (\bibinfo{year}{2020}).

\bibitem{Kong2021}
\bibinfo{author}{Kong, L.} \emph{et~al.}
\newblock \bibinfo{title}{Majorana zero modes in impurity-assisted vortex of
  LiFeAs superconductor}.
\newblock \emph{\bibinfo{journal}{Nature Communications}}
  \textbf{\bibinfo{volume}{12}}, \bibinfo{pages}{1--11} (\bibinfo{year}{2021}).

\bibitem{Li2022}
\bibinfo{author}{Li, M.} \emph{et~al.}
\newblock \bibinfo{title}{Ordered and tunable Majorana-zero-mode lattice in
  naturally strained LiFeAs}.
\newblock \emph{\bibinfo{journal}{Nature}} \bibinfo{pages}{1--6}
  (\bibinfo{year}{2022}).

\bibitem{Yuan2019}
\bibinfo{author}{Yuan, Y.} \emph{et~al.}
\newblock \bibinfo{title}{Evidence of anisotropic Majorana bound states in
  2M-WS$_2$}.
\newblock \emph{\bibinfo{journal}{Nature Physics}}
  \textbf{\bibinfo{volume}{15}}, \bibinfo{pages}{1046--1051}
  (\bibinfo{year}{2019}).

\bibitem{Fang2019}
\bibinfo{author}{Fang, Y.} \emph{et~al.}
\newblock \bibinfo{title}{Discovery of superconductivity in 2M WS$_2$ with
  possible topological surface states}.
\newblock \emph{\bibinfo{journal}{Advanced Materials}}
  \textbf{\bibinfo{volume}{31}}, \bibinfo{pages}{1901942}
  (\bibinfo{year}{2019}).

\bibitem{Li2021}
\bibinfo{author}{Li, Y.} \emph{et~al.}
\newblock \bibinfo{title}{Observation of topological superconductivity in a
  stoichiometric transition metal dichalcogenide 2M-WS$_2$}.
\newblock \emph{\bibinfo{journal}{Nature communications}}
  \textbf{\bibinfo{volume}{12}}, \bibinfo{pages}{1--7} (\bibinfo{year}{2021}).

\bibitem{Lv2017}
\bibinfo{author}{Lv, Y.-F.} \emph{et~al.}
\newblock \bibinfo{title}{Experimental signature of topological
  superconductivity and Majorana zero modes on $\beta$-Bi$_2$Pd thin films}.
\newblock \emph{\bibinfo{journal}{Science bulletin}}
  \textbf{\bibinfo{volume}{62}}, \bibinfo{pages}{852--856}
  (\bibinfo{year}{2017}).

\bibitem{Guan2019}
\bibinfo{author}{Guan, J.-Y.} \emph{et~al.}
\newblock \bibinfo{title}{Experimental evidence of anomalously large
  superconducting gap on topological surface state of $\beta$-Bi$_2$Pd film}.
\newblock \emph{\bibinfo{journal}{Science Bulletin}}
  \textbf{\bibinfo{volume}{64}}, \bibinfo{pages}{1215--1221}
  (\bibinfo{year}{2019}).

\bibitem{Li2019}
\bibinfo{author}{Li, Y.}, \bibinfo{author}{Xu, X.}, \bibinfo{author}{Lee,
  M.-H.}, \bibinfo{author}{Chu, M.-W.} \& \bibinfo{author}{Chien, C.}
\newblock \bibinfo{title}{Observation of half-quantum flux in the
  unconventional superconductor $\beta$-Bi$_2$Pd}.
\newblock \emph{\bibinfo{journal}{Science}} \textbf{\bibinfo{volume}{366}},
  \bibinfo{pages}{238--241} (\bibinfo{year}{2019}).

\bibitem{Wang2020}
\bibinfo{author}{Wang, Z.} \emph{et~al.}
\newblock \bibinfo{title}{Evidence for dispersing 1D Majorana channels in an
  iron-based superconductor}.
\newblock \emph{\bibinfo{journal}{Science}} \textbf{\bibinfo{volume}{367}},
  \bibinfo{pages}{104--108} (\bibinfo{year}{2020}).

\bibitem{Qi2010}
\bibinfo{author}{Qi, X.-L.}, \bibinfo{author}{Hughes, T.~L.} \&
  \bibinfo{author}{Zhang, S.-C.}
\newblock \bibinfo{title}{Chiral topological superconductor from the quantum
  Hall state}.
\newblock \emph{\bibinfo{journal}{Phys. Rev. B}} \textbf{\bibinfo{volume}{82}},
  \bibinfo{pages}{184516} (\bibinfo{year}{2010}).

\bibitem{Wang2015}
\bibinfo{author}{Wang, J.}, \bibinfo{author}{Zhou, Q.}, \bibinfo{author}{Lian,
  B.} \& \bibinfo{author}{Zhang, S.-C.}
\newblock \bibinfo{title}{Chiral topological superconductor and half-integer
  conductance plateau from quantum anomalous Hall plateau transition}.
\newblock \emph{\bibinfo{journal}{Phys. Rev. B}} \textbf{\bibinfo{volume}{92}},
  \bibinfo{pages}{064520} (\bibinfo{year}{2015}).

\bibitem{He2019}
\bibinfo{author}{He, J.~J.}, \bibinfo{author}{Liang, T.},
  \bibinfo{author}{Tanaka, Y.} \& \bibinfo{author}{Nagaosa, N.}
\newblock \bibinfo{title}{Platform of chiral Majorana edge modes and its
  quantum transport phenomena}.
\newblock \emph{\bibinfo{journal}{Communications Physics}}
  \textbf{\bibinfo{volume}{2}}, \bibinfo{pages}{1--7} (\bibinfo{year}{2019}).

\bibitem{Zhang2021a}
\bibinfo{author}{Zhang, X.} \& \bibinfo{author}{Liu, F.}
\newblock \bibinfo{title}{Prediction of Majorana edge states from magnetized
  topological surface states}.
\newblock \emph{\bibinfo{journal}{Phys. Rev. B}}
  \textbf{\bibinfo{volume}{103}}, \bibinfo{pages}{024405}
  (\bibinfo{year}{2021}).

\bibitem{Zhang2021b}
\bibinfo{author}{Zhang, X.} \emph{et~al.}
\newblock \bibinfo{title}{Prediction of intrinsic topological superconductivity
  in Mn-doped GeTe monolayer from first-principles}.
\newblock \emph{\bibinfo{journal}{npj Computational Materials}}
  \textbf{\bibinfo{volume}{7}}, \bibinfo{pages}{1--8} (\bibinfo{year}{2021}).

\bibitem{Menard2017}
\bibinfo{author}{M{\'e}nard, G.~C.} \emph{et~al.}
\newblock \bibinfo{title}{Two-dimensional topological superconductivity in
  Pb/Co/Si(111)}.
\newblock \emph{\bibinfo{journal}{Nature Communications}}
  \textbf{\bibinfo{volume}{8}}, \bibinfo{pages}{1--7} (\bibinfo{year}{2017}).

\bibitem{Alexandra2019}
\bibinfo{author}{Palacio-Morales, A.} \emph{et~al.}
\newblock \bibinfo{title}{Atomic-scale interface engineering of Majorana edge
  modes in a 2D magnet-superconductor hybrid system}.
\newblock \emph{\bibinfo{journal}{Science Advances}}
  \textbf{\bibinfo{volume}{5}}, \bibinfo{pages}{eaav6600}
  (\bibinfo{year}{2019}).

\bibitem{Kezilebieke2020}
\bibinfo{author}{Kezilebieke, S.} \emph{et~al.}
\newblock \bibinfo{title}{Topological superconductivity in a van der Waals
  heterostructure}.
\newblock \emph{\bibinfo{journal}{Nature}} \textbf{\bibinfo{volume}{588}},
  \bibinfo{pages}{424--428} (\bibinfo{year}{2020}).

\bibitem{Biao2018}
\bibinfo{author}{Lian, B.}, \bibinfo{author}{Sun, X.-Q.},
  \bibinfo{author}{Vaezi, A.}, \bibinfo{author}{Qi, X.-L.} \&
  \bibinfo{author}{Zhang, S.-C.}
\newblock \bibinfo{title}{Topological quantum computation based on chiral
  Majorana fermions}.
\newblock \emph{\bibinfo{journal}{Proceedings of the National Academy of
  Sciences}} \textbf{\bibinfo{volume}{115}}, \bibinfo{pages}{10938--10942}
  (\bibinfo{year}{2018}).

\bibitem{Huang2021}
\bibinfo{author}{Huang, K.} \emph{et~al.}
\newblock \bibinfo{title}{Observation of topological Dirac fermions and surface
  states in superconducting $\mathrm{Ba}{\mathrm{Sn}}_{3}$}.
\newblock \emph{\bibinfo{journal}{Phys. Rev. B}}
  \textbf{\bibinfo{volume}{103}}, \bibinfo{pages}{155148}
  (\bibinfo{year}{2021}).

\bibitem{Chen2020}
\bibinfo{author}{Chen, C.} \emph{et~al.}
\newblock \bibinfo{title}{Observation of topological electronic structure in
  quasi-1D superconductor TaSe$_3$}.
\newblock \emph{\bibinfo{journal}{Matter}} \textbf{\bibinfo{volume}{3}},
  \bibinfo{pages}{2055--2065} (\bibinfo{year}{2020}).

\bibitem{Ortiz2020}
\bibinfo{author}{Ortiz, B.~R.} \emph{et~al.}
\newblock \bibinfo{title}{$\mathrm{Cs}{\mathrm{V}}_{3}{\mathrm{Sb}}_{5}$: A
  ${\mathbb{Z}}_{2}$ topological kagome metal with a superconducting ground
  state}.
\newblock \emph{\bibinfo{journal}{Phys. Rev. Lett.}}
  \textbf{\bibinfo{volume}{125}}, \bibinfo{pages}{247002}
  (\bibinfo{year}{2020}).

\bibitem{Leng2017}
\bibinfo{author}{Leng, H.}, \bibinfo{author}{Paulsen, C.},
  \bibinfo{author}{Huang, Y.~K.} \& \bibinfo{author}{de~Visser, A.}
\newblock \bibinfo{title}{Type-I superconductivity in the Dirac semimetal
  ${\mathrm{PdTe}}_{2}$}.
\newblock \emph{\bibinfo{journal}{Phys. Rev. B}} \textbf{\bibinfo{volume}{96}},
  \bibinfo{pages}{220506} (\bibinfo{year}{2017}).

\bibitem{Das2018}
\bibinfo{author}{Das, S.} \emph{et~al.}
\newblock \bibinfo{title}{Conventional superconductivity in the type-II Dirac
  semimetal ${\mathrm{PdTe}}_{2}$}.
\newblock \emph{\bibinfo{journal}{Phys. Rev. B}} \textbf{\bibinfo{volume}{97}},
  \bibinfo{pages}{014523} (\bibinfo{year}{2018}).

\bibitem{Kudo2016}
\bibinfo{author}{Kudo, K.}, \bibinfo{author}{Ishii, H.} \&
  \bibinfo{author}{Nohara, M.}
\newblock \bibinfo{title}{Composition-induced structural instability and
  strong-coupling superconductivity in
  ${\mathrm{Au}}_{1\ensuremath{-}x}{\mathrm{Pd}}_{x}{\mathrm{Te}}_{2}$}.
\newblock \emph{\bibinfo{journal}{Phys. Rev. B}} \textbf{\bibinfo{volume}{93}},
  \bibinfo{pages}{140505} (\bibinfo{year}{2016}).

\bibitem{Lee2013}
\bibinfo{author}{Lee, D.~S.} \emph{et~al.}
\newblock \bibinfo{title}{Crystal structure, properties and nanostructuring of
  a new layered chalcogenide semiconductor, Bi$_2$MnTe$_4$}.
\newblock \emph{\bibinfo{journal}{CrystEngComm}} \textbf{\bibinfo{volume}{15}},
  \bibinfo{pages}{5532--5538} (\bibinfo{year}{2013}).

\bibitem{Zhang2019}
\bibinfo{author}{Zhang, D.} \emph{et~al.}
\newblock \bibinfo{title}{Topological axion states in the magnetic insulator
  ${\mathrm{MnBi}}_{2}{\mathrm{Te}}_{4}$ with the quantized magnetoelectric
  effect}.
\newblock \emph{\bibinfo{journal}{Phys. Rev. Lett.}}
  \textbf{\bibinfo{volume}{122}}, \bibinfo{pages}{206401}
  (\bibinfo{year}{2019}).

\bibitem{Wu2018}
\bibinfo{author}{Wu, Q.}, \bibinfo{author}{Zhang, S.}, \bibinfo{author}{Song,
  H.-F.}, \bibinfo{author}{Troyer, M.} \& \bibinfo{author}{Soluyanov, A.~A.}
\newblock \bibinfo{title}{Wanniertools: An open-source software package for
  novel topological materials}.
\newblock \emph{\bibinfo{journal}{Computer Physics Communications}}
  \textbf{\bibinfo{volume}{224}}, \bibinfo{pages}{405--416}
  (\bibinfo{year}{2018}).

\bibitem{Yu2011}
\bibinfo{author}{Yu, R.}, \bibinfo{author}{Qi, X.~L.},
  \bibinfo{author}{Bernevig, A.}, \bibinfo{author}{Fang, Z.} \&
  \bibinfo{author}{Dai, X.}
\newblock \bibinfo{title}{Equivalent expression of ${\mathbb{Z}}_{2}$
  topological invariant for band insulators using the non-abelian berry
  connection}.
\newblock \emph{\bibinfo{journal}{Phys. Rev. B}} \textbf{\bibinfo{volume}{84}},
  \bibinfo{pages}{075119} (\bibinfo{year}{2011}).

\bibitem{Soluyanov2011}
\bibinfo{author}{Soluyanov, A.~A.} \& \bibinfo{author}{Vanderbilt, D.}
\newblock \bibinfo{title}{Computing topological invariants without inversion
  symmetry}.
\newblock \emph{\bibinfo{journal}{Phys. Rev. B}} \textbf{\bibinfo{volume}{83}},
  \bibinfo{pages}{235401} (\bibinfo{year}{2011}).

\bibitem{Gresch2017}
\bibinfo{author}{Gresch, D.} \emph{et~al.}
\newblock \bibinfo{title}{Z$_2$pack: Numerical implementation of hybrid wannier
  centers for identifying topological materials}.
\newblock \emph{\bibinfo{journal}{Phys. Rev. B}} \textbf{\bibinfo{volume}{95}},
  \bibinfo{pages}{075146} (\bibinfo{year}{2017}).

\bibitem{Zhang2009}
\bibinfo{author}{Zhang, H.} \emph{et~al.}
\newblock \bibinfo{title}{Topological insulators in Bi$_2$Se$_3$, Bi$_2$Te$_3$ and Sb$_2$Te$_3$
  with a single Dirac cone on the surface}.
\newblock \emph{\bibinfo{journal}{Nature Physics}}
  \textbf{\bibinfo{volume}{5}}, \bibinfo{pages}{438--442}
  (\bibinfo{year}{2009}).

\bibitem{Chen2009}
\bibinfo{author}{Chen, Y.} \emph{et~al.}
\newblock \bibinfo{title}{Experimental realization of a three-dimensional
  topological insulator, Bi$_2$Te$_3$}.
\newblock \emph{\bibinfo{journal}{Science}} \textbf{\bibinfo{volume}{325}},
  \bibinfo{pages}{178--181} (\bibinfo{year}{2009}).

\bibitem{Matthias1953}
\bibinfo{author}{Matthias, B.~T.}
\newblock \bibinfo{title}{Superconducting compounds of nonsuperconducting
  elements}.
\newblock \emph{\bibinfo{journal}{Phys. Rev.}} \textbf{\bibinfo{volume}{90}},
  \bibinfo{pages}{487--487} (\bibinfo{year}{1953}).

\bibitem{Karki2012}
\bibinfo{author}{Karki, A.~B.}, \bibinfo{author}{Browne, D.~A.},
  \bibinfo{author}{Stadler, S.}, \bibinfo{author}{Li, J.} \&
  \bibinfo{author}{Jin, R.}
\newblock \bibinfo{title}{PdTe: a strongly coupled superconductor}.
\newblock \emph{\bibinfo{journal}{Journal of Physics: Condensed Matter}}
  \textbf{\bibinfo{volume}{24}}, \bibinfo{pages}{055701}
  (\bibinfo{year}{2012}).

\bibitem{SM}
\emph{\bibinfo{journal}{See Supplemental Materials for more details}.}

\bibitem{Sharma2020}
\bibinfo{author}{Sharma, M.}, \bibinfo{author}{Sang, L.},
  \bibinfo{author}{Rani, P.}, \bibinfo{author}{Wang, X.} \&
  \bibinfo{author}{Awana, V.}
\newblock \bibinfo{title}{Bulk superconductivity below 6 K in PdBi$_2$Te$_3$
  topological single crystal}.
\newblock \emph{\bibinfo{journal}{Journal of Superconductivity and Novel
  Magnetism}} \textbf{\bibinfo{volume}{33}}, \bibinfo{pages}{1243--1247}
  (\bibinfo{year}{2020}).

\bibitem{Wang2021}
\bibinfo{author}{Wang, X.} \emph{et~al.}
\newblock \bibinfo{title}{Identify the nematic superconductivity of topological
  superconductor Pd$_x$Bi$_2$Te$_3$ by angle-dependent upper critical field
  measurement}.
\newblock \emph{\bibinfo{journal}{Journal of Superconductivity and Novel
  Magnetism}} \textbf{\bibinfo{volume}{34}}, \bibinfo{pages}{3045--3052}
  (\bibinfo{year}{2021}).

\bibitem{Fu2007}
\bibinfo{author}{Fu, L.}, \bibinfo{author}{Kane, C.~L.} \&
  \bibinfo{author}{Mele, E.~J.}
\newblock \bibinfo{title}{Topological insulators in three dimensions}.
\newblock \emph{\bibinfo{journal}{Phys. Rev. Lett.}}
  \textbf{\bibinfo{volume}{98}}, \bibinfo{pages}{106803}
  (\bibinfo{year}{2007}).

\bibitem{Baroni2001}
\bibinfo{author}{Baroni, S.}, \bibinfo{author}{de~Gironcoli, S.},
  \bibinfo{author}{Dal~Corso, A.} \& \bibinfo{author}{Giannozzi, P.}
\newblock \bibinfo{title}{Phonons and related crystal properties from
  density-functional perturbation theory}.
\newblock \emph{\bibinfo{journal}{Rev. Mod. Phys.}}
  \textbf{\bibinfo{volume}{73}}, \bibinfo{pages}{515--562}
  (\bibinfo{year}{2001}).

\bibitem{McMillan1968}
\bibinfo{author}{McMillan, W.~L.}
\newblock \bibinfo{title}{Transition temperature of strong-coupled
  superconductors}.
\newblock \emph{\bibinfo{journal}{Phys. Rev.}} \textbf{\bibinfo{volume}{167}},
  \bibinfo{pages}{331--344} (\bibinfo{year}{1968}).

\bibitem{Allen1975}
\bibinfo{author}{Allen, P.~B.} \& \bibinfo{author}{Dynes, R.~C.}
\newblock \bibinfo{title}{Transition temperature of strong-coupled
  superconductors reanalyzed}.
\newblock \emph{\bibinfo{journal}{Phys. Rev. B}} \textbf{\bibinfo{volume}{12}},
  \bibinfo{pages}{905--922} (\bibinfo{year}{1975}).

\bibitem{Kresse1993}
\bibinfo{author}{Kresse, G.} \& \bibinfo{author}{Hafner, J.}
\newblock \bibinfo{title}{Ab initio molecular dynamics for open-shell
  transition metals}.
\newblock \emph{\bibinfo{journal}{Physical Review B}}
  \textbf{\bibinfo{volume}{48}}, \bibinfo{pages}{13115} (\bibinfo{year}{1993}).

\bibitem{Kresse1996}
\bibinfo{author}{Kresse, G.} \& \bibinfo{author}{Furthm{\"u}ller, J.}
\newblock \bibinfo{title}{Efficiency of ab-initio total energy calculations for
  metals and semiconductors using a plane-wave basis set}.
\newblock \emph{\bibinfo{journal}{Computational Materials Science}}
  \textbf{\bibinfo{volume}{6}}, \bibinfo{pages}{15--50} (\bibinfo{year}{1996}).

\bibitem{Perdew1996}
\bibinfo{author}{Perdew, J.~P.}, \bibinfo{author}{Burke, K.} \&
  \bibinfo{author}{Ernzerhof, M.}
\newblock \bibinfo{title}{Generalized gradient approximation made simple}.
\newblock \emph{\bibinfo{journal}{Phys. Rev. Lett.}}
  \textbf{\bibinfo{volume}{77}}, \bibinfo{pages}{3865--3868}
  (\bibinfo{year}{1996}).

\bibitem{Klime2011}
\bibinfo{author}{Klime\ifmmode~\check{s}\else \v{s}\fi{}, J. c.~v.},
  \bibinfo{author}{Bowler, D.~R.} \& \bibinfo{author}{Michaelides, A.}
\newblock \bibinfo{title}{Van der waals density functionals applied to solids}.
\newblock \emph{\bibinfo{journal}{Phys. Rev. B}} \textbf{\bibinfo{volume}{83}},
  \bibinfo{pages}{195131} (\bibinfo{year}{2011}).

\bibitem{Giannozzi2009}
\bibinfo{author}{Giannozzi, P.} \emph{et~al.}
\newblock \bibinfo{title}{Quantum espresso: a modular and open-source software
  project for quantum simulations of materials}.
\newblock \emph{\bibinfo{journal}{Journal of physics: Condensed Matter}}
  \textbf{\bibinfo{volume}{21}}, \bibinfo{pages}{395502}
  (\bibinfo{year}{2009}).

\bibitem{Togo2015}
\bibinfo{author}{Togo, A.} \& \bibinfo{author}{Tanaka, I.}
\newblock \bibinfo{title}{First principles phonon calculations in materials
  science}.
\newblock \emph{\bibinfo{journal}{Scr. Mater.}} \textbf{\bibinfo{volume}{108}},
  \bibinfo{pages}{1--5} (\bibinfo{year}{2015}).

\bibitem{Mostofi2008}
\bibinfo{author}{Mostofi, A.~A.} \emph{et~al.}
\newblock \bibinfo{title}{wannier90: A tool for obtaining maximally-localised
  wannier functions}.
\newblock \emph{\bibinfo{journal}{Computer Physics Communications}}
  \textbf{\bibinfo{volume}{178}}, \bibinfo{pages}{685--699}
  (\bibinfo{year}{2008}).

\end{thebibliography}
\end{document}

\title{\emph{Supplementary Material}: Topological superconductor candidates PdBi$_2$Te$_4$ and PdBi$_2$Te$_5$ from a generic ab initio strategy}

\author{Aiyun Luo}
\thanks{These authors made equal contributions to this work.}
\affiliation{Wuhan National High Magnetic Field Center $\&$ School of Physics, Huazhong University of Science and Technology, Wuhan 430074, China}
\author{Ying Li}
\thanks{These authors made equal contributions to this work.}
\affiliation{Wuhan National High Magnetic Field Center $\&$ School of Physics, Huazhong University of Science and Technology, Wuhan 430074, China}
\author{Yi Qin}
\thanks{These authors made equal contributions to this work.}
\affiliation{Wuhan National High Magnetic Field Center $\&$ School of Physics, Huazhong University of Science and Technology, Wuhan 430074, China}
\author{Jingnan Hu}
\affiliation{Wuhan National High Magnetic Field Center $\&$ School of Physics, Huazhong University of Science and Technology, Wuhan 430074, China}
\author{Biao Lian}
\affiliation{Department of Physics, Princeton University, Princeton, NJ 08544, United States of America}
\author{Gang Xu}
\email[e-mail address: ]{gangxu@hust.edu.cn}
\affiliation{Wuhan National High Magnetic Field Center $\&$ School of Physics, Huazhong University of Science and Technology, Wuhan 430074, China}
\affiliation{Institute for Quantum Science and Engineering, Huazhong University of Science and Technology, Wuhan 430074, China}
\affiliation{Wuhan Institute of Quantum Technology, Wuhan 430074, China}

\maketitle

\section{The crystal structures of PdBi$_2$Te$_5$ and PdBi$_2$Te$_4$}
\begin{figure}[!h]
  \includegraphics[width=0.8\columnwidth]{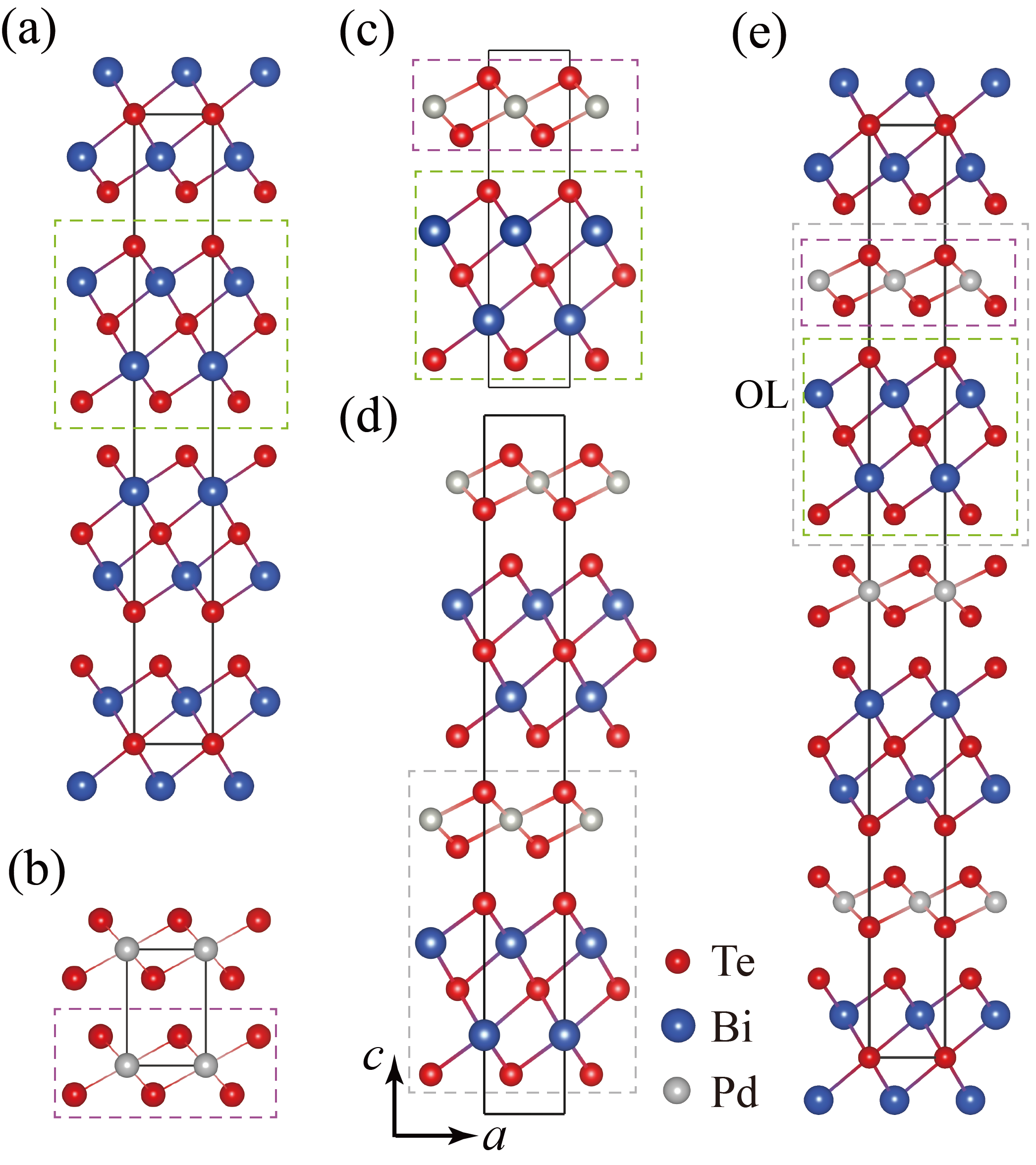}
  \caption{The crystal structures of different stacking sequences of PdBi$_2$Te$_5$.
  (a) A side view of the crystal structures of Bi$_2$Te$_3$, the quintuple-layer block is marked by the green dashed rectangle.
  (b) PdTe$_2$, the triple-layer block is marked by the purple dashed rectangle.
  PdBi$_2$Te$_5$ for (c) AA stacking, (d) AB stacking, and (e) ABC stacking, respectively.
  A octuple-layer(OL) block of PdBi$_2$Te$_5$ is marked by the grey dashed rectangle.}
  \label{fig:S1}
\end{figure}

The crystal structures of PdBi$_2$Te$_5$ are shown in Fig.~\ref{fig:S1}.
Since both Bi$_2$Te$_3$(Fig.~\ref{fig:S1}(a)) and PdTe$_2$(Fig.~\ref{fig:S1}(b)) are Van der Waals materials,
a naturally stable unit of PdBi$_2$Te$_5$ would be an octuple-layer (OL) block as shown in Fig.~\ref{fig:S1}(e).
With different stacking sequences along $c$-direction,
the crystal structures of AA-stacking, AB-stacking and ABC-stacking are shown in Fig.~\ref{fig:S1}(c), (d) and (e), respectively.
As a result, the relaxed lattice parameters and total energies of these structures
are tabulated in Table.~\ref{tab:S1}.
Our calculated results demonstrate that the most stable structure of PdBi$_2$Te$_5$ is the ABC-stacking rhombohedral unit cell as shown in Fig.~\ref{fig:S1}(e),
which is 73~meV/f.u. and 46~meV/f.u. lower than the AA and AB stacking structures.

\begin{figure}[!h]
  \includegraphics[width=0.8\columnwidth]{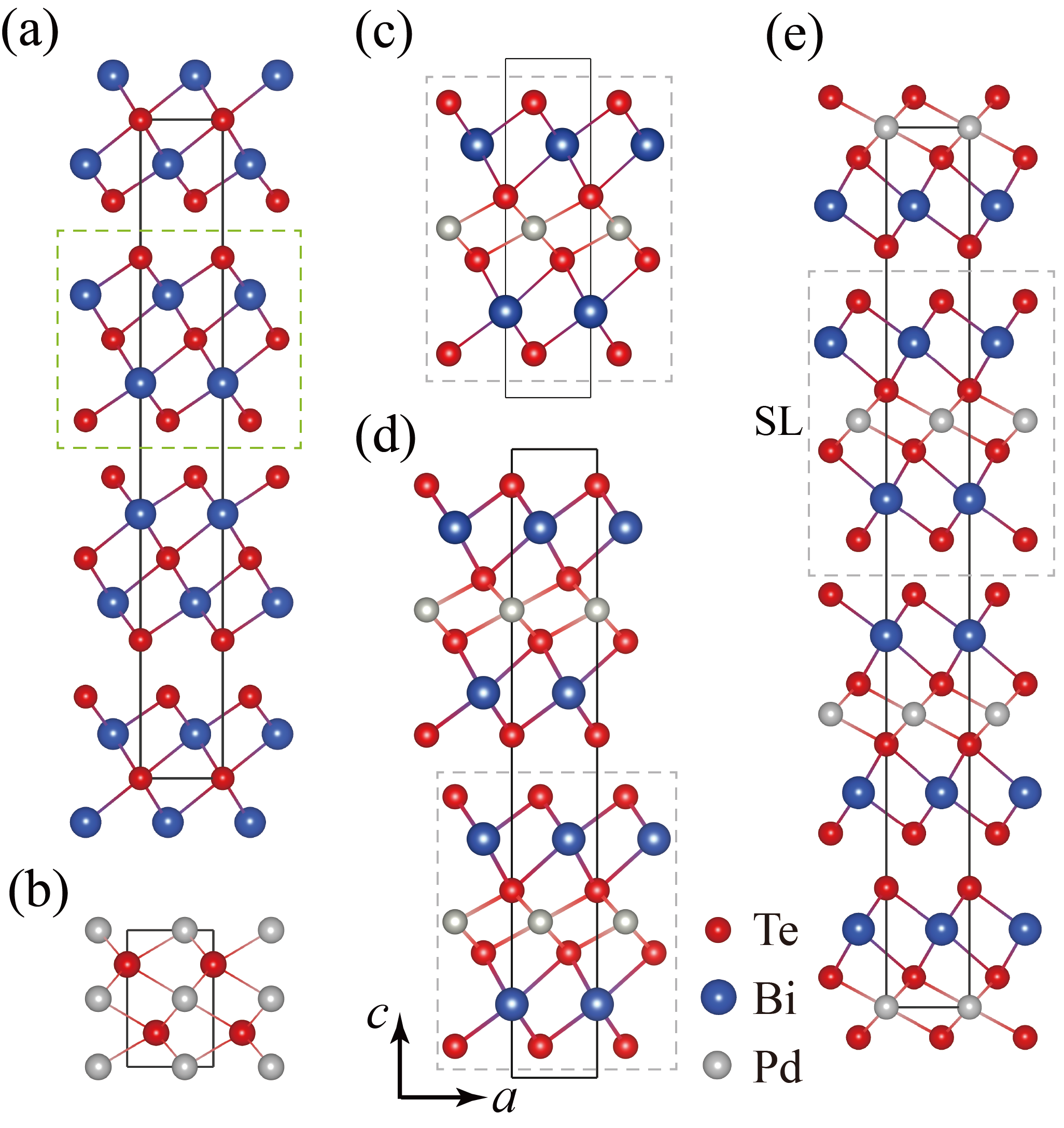}
  \caption{The crystal structures of different stacking sequences of PdBi$_2$Te$_4$.
  (a) Bi$_2$Te$_3$. (b) PdTe.
  PdBi$_2$Te$_4$ for (c) AA stacking, (d) AB stacking, and (e) ABC stacking, respectively.
  A SL block of PdBi$_2$Te$_4$ is marked by the grey dashed rectangle.}
  \label{fig:S2}
\end{figure}

The crystal structures of PdBi$_2$Te$_4$ are shown in Fig.~\ref{fig:S2}, which is formed by
the integration of Bi$_2$Te$_3$(Fig.~\ref{fig:S2}(a)) and PdTe(Fig.~\ref{fig:S2}(b)).
Like the construction of MnBi$_2$Te$_4$~\cite{Lee2013, Zhang2019},
a naturally stable unit of PdBi$_2$Te$_4$ would be a septuple-layer (SL) block
as shown by the grey dash rectangle in Fig.~\ref{fig:S2}(c)-(e).
Similar to the case of PdBi$_2$Te$_5$, the relaxed lattice parameters and total energies of
AA-stacking, AB-stacking, and ABC-stacking sequences of PdBi$_2$Te$_4$
are tabulated in Table.~\ref{tab:S2}.
Our calculations demonstrate that SL-PdBi$_2$Te$_4$ favor the ABC stacking rhombohedral unit cell as shown in Fig.~\ref{fig:S2}(e),
which is 73~meV/f.u. and 12~meV/f.u. lower than the AA(Fig.~\ref{fig:S2}(c)) and AB(Fig.~\ref{fig:S2}(d)) stacking structures.

\begin{table*}[!h]
\centering
\setlength{\tabcolsep}{3mm}
\caption{The crystal parameters and total energies of different stacking sequences of PdBi$_2$Te$_5$ from our relaxed results.}
\begin{tabular}{cccccccc}
\hline
\hline
Compound & Stacking & Space group & Lattice (\AA) & Atom & Wyckoff & Coordinate & Energy(eV/f.u.) \\
\hline
		PdBi$_{2}$Te$_{5}$ & AA &$P3m1$ & a = b = 4.287 & Pd & $1c$ & (0.667, 0.333, 0.013) & -30.385\\
		& & & c = 16.325 & Bi1 & $1a$ & 0, 0, 0.363 & \\
		& & & & Bi2 & $1a$ & 0, 0, 0.623 &\\
		& & & & Te1 & $1b$ & (0.333, 0.667, 0.090) &\\
		& & & & Te2 & $1b$ & (0.333, 0.667, 0.492) &\\
		& & & & Te3 & $1c$ & (0.667, 0.333, 0.251) &\\
		& & & & Te4 & $1c$ & (0.667, 0.333, 0.936) &\\
		& & & & Te5 & $1a$ & (0, 0, 0.733)&\\
\hline
		PdBi$_{2}$Te$_{5}$ & AB &$P3m1$ & a = b = 4.315 & Pd1 & $1c$ & (0.667, 0.333, 0.483) & -30.412\\
		& & & c = 30.987 & Pd2 & $1a$ & (0, 0, 0) & \\
		& & & & Bi1 & $1b$ & (0.333, 0.667, 0.792) &\\
		& & & & Bi2 & $1c$ & (0.667, 0.333, 0.658) &\\
		& & & & Bi3 & $1c$ & (0.667, 0.333, 0.308) &\\
		& & & & Bi4 & $1a$ & (0, 0, 0.175) &\\
		& & & & Te1 & $1b$ & (0.333, 0.667, 0.046) &\\
		& & & & Te2 & $1b$ & (0.333, 0.667, 0.437) &\\
		& & & & Te3 & $1b$ & (0.333, 0.667, 0.600) &\\
		& & & & Te4 & $1b$ & (0.333, 0.667, 0.241) &\\
		& & & & Te5 & $1c$ & (0.667, 0.333, 0.956) &\\
		& & & & Te6 & $1c$ & (0.667, 0.333, 0.115) &\\
		& & & & Te7 & $1c$ & (0.667, 0.333, 0.851) &\\
		& & & & Te8 & $1a$ & (0, 0, 0.528) &\\
		& & & & Te9 & $1a$ & (0, 0, 0.726) &\\
		& & & & Te10 & $1a$ & (0, 0, 0.367) &\\
\hline
		PdBi$_{2}$Te$_{5}$ & ABC &$R\bar{3}m$ & a = b = 4.304 & Pd & $3b$ & (0, 0, 0.5)& -30.458\\
		& & & c = 46.391 & Bi & $6c$ & (0, 0, 0.3788)\\
		& & & & Te1 & $3a$ &(0, 0, 0)\\
		& & & & Te2 & $6c$ &(0, 0, 0.1396)\\
		& & & & Te3 & $6c$ &(0, 0, 0.2488)\\
\hline
\hline
\end{tabular}
\label{tab:S1}
\end{table*}

\begin{table*}[!h]
\centering
\setlength{\tabcolsep}{3mm}
\caption{The crystal parameters and total energies of different stacking sequences of PdBi$_2$Te$_4$ from our relaxed results.}
\begin{tabular}{cccccccc}
\hline
\hline
Compound & Stacking & Space group & Lattice (\AA) & Atom & Wyckoff & Coordinate & Energy(eV/f.u.) \\
\hline
        PdBi$_{2}$Te$_{4}$ & AA & $P\bar{3}m\rm{1}$ & a = b = 4.275 & Pd & $1b$ & (0, 0, 0.5) & -26.776\\
		& & & c = 13.035 & Bi & $2d$ & (0.667, 0.333, 0.2325) \\
		& & & & Te1 & $2d$ & (0.333, 0.667, 0.397)\\
		& & & & Te2 & $2c$ & (0, 0, 0.09)\\
\hline
        PdBi$_{2}$Te$_{4}$ & AB & $P\bar{3}m\rm{1}$ & a = b = 4.275 & Pd & $2d$ & (0.333, 0.667, 0.75025)& -26.837\\
		& & & c = 26.070 & Bi1 & $2c$ & (0, 0, 0.61675)\\
		& & & & Bi2 & $2d$ & (0.667, 0.333, 0.88375)\\
		& & & & Te1 & $2d$ & (0.333, 0.667, 0.95375)\\
		& & & & Te2 & $2d$ & (0.333, 0.667, 0.54675)\\
		& & & & Te3 & $2c$ & (0, 0, 0.80175)\\
		& & & & Te4 & $2d$ & (0.667, 0.333, 0.88375)\\
\hline
		PdBi$_{2}$Te$_{4}$ & ABC & $R\bar{3}m$  & a = b =4.296 & Pd & $3a$ & (0, 0, 0) & -26.849\\
		& & & c = 40.546 & Bi & $6c$ &(0, 0, 0.4215) \\
		& & & & Te1 & $6c$ &(0, 0, 0.3000)\\
		& & & & Te2 & $6c$ &(0, 0, 0.8668)\\
\hline
\hline
\end{tabular}
\label{tab:S2}
\end{table*}

\section{The detailed crystal parameters and formation energies of Pd-Bi-Te compounds for convex hull}
The fundamental thermodynamic nature of Pd-Bi-Te compounds can be fully characterized in terms of the convex hull diagram~\cite{Sun2019,Sharan2021}, 
which is defined as formation energy $versus$ composition diagram with the ground state at special compositions.
For all of the known Pd-Bi-Te compounds as tabulated in Table.~\ref{tab:convex-hull-struc},
we calculate their formation energy and tabular the results in Table.~\ref{tab:convex-hull}.
Based on these results, we construct the convex hull diagram of Pd-Bi-Te compounds in Fig.3(b) of the main text.
The ternary compounds PdBi$_2$Te$_3$, PdBi$_2$Te$_4$ and Pd$_2$Te$_5$
are 65~meV/atom, 59~meV/atom and 13~meV/atom above the convex hull respectively, indicating that they are metastable phases with respect to decomposition into the energetically favorable binary phases.
Significantly, PdBi$_2$Te$_3$ above the convex hull has already been synthesized in experiments,
which is higher 52~meV/atom and 7~meV/atom than PdBi$_2$Te$_5$ and PdBi$_2$Te$_4$.
Therefore, we expect that PdBi$_2$Te$_5$ and PdBi$_2$Te$_4$ could be synthesized in the future.

\begin{table*}[!h]
	\centering
	\setlength{\tabcolsep}{4mm}
	\caption{The space group, lattice constants and atomic coordinates of Pd-Bi-Te compounds used in the convex hull.}
	\begin{tabular}{cccccc}
		\hline
        \hline
		Compound & Space group & Lattice (\AA) & Atom & Wyckoff & Coordinate \\
		\hline
		Pd & $Fm\bar{3}m$  & a = b = c = 3.889 & Pd & $4a$ & (0, 0, 0)\\
  	    Bi & $R\bar{3}m$ & a = 4.523, c = 11.800 & Bi & $3b$ & (0, 0, 0.227)\\
        Te & $P$3$_1$21 & a = 4.445, c = 5.91  & Te & $3a$ & (0.225, 0.0, 0.0)\\
		PdBi & $Cmc\rm{2_1}$ & a = 8.707 & Pd1 & $8b$ & (0.274, 0.125, 0.053)\\
		& & b = 7.203 & Pd2 & $4a$ & (0, 0.108, 0.225)\\
		& & c = 10.662 & Pd3 & $4a$ & (0, 0.65, 0.225)\\
		& & & Bi1 & $8b$ & (0.226, 0.375, 0.278)\\
		& & & Bi2 & $4a$ & (0, 0.108, 0.5)\\
		& & & Bi3 & $4a$ & (0, 0.35, 0)\\
		PdBi$_2$ & $I\rm{4}/mmm$ & a = b = 3.362 & Pd & $2a$ & (0, 0, 0)\\
		& & c = 12.983 & Bi & $4e$ & (0, 0, 0.363)\\
		PdTe & $P\rm{6_3}/mmc$ & a = b = 4.152 & Pd & $2a$& (0, 0, 0)\\
		& & c = 5.671 & Te & $2c$ & (0.333, 0.667, 0.25)\\
		PdTe$_2$ & $P\bar{3}m\rm{1}$ & a = b = 4.028 & Pd & $1a$ & (0, 0, 0)\\
		& & c = 5.118 & Te & $2d$ & (0.333, 0.667, 0.25)\\
		BiTe & $P\bar{3}m\rm{1}$ & a = b = 4.400 & Bi1 & $2d$ & (0.333, 0.667, 0.291)\\
		& & c = 24.000 & Bi2 & $2d$ & (0.333, 0.667, 0.541)\\
		& & & Bi3 & $2c$ & (0, 0, 0.126)\\
		& & & Te1 & $2d$ & (0.333, 0.667, 0.056)\\
		& & & Te2 & $2d$ & (0.333, 0.667, 0.789)\\
		& & & Te3 & $2c$ & (0, 0, 0.362)\\
		Bi$_2$Te$_3$ & $R\bar{3}m$ & a = b = 4.35 & Bi & $6c$ & (0, 0, 0.600)\\
		& & c = 30.36 & Te1 & $6c$ & (0, 0, 0.790)\\
		& & & Te2 & $3a$ & (0, 0, 0)\\
		Bi$_4$Te$_3$ & $R\bar{3}m$ & a = b = 4.451 & Bi1 & $6c$ & (0, 0, 0.146)\\
		& & c = 41.888 & Bi2 & $6c$ & (0, 0, 0.283)\\
		& & & Te1 & $6c$ & (0, 0, 0.426)\\
		& & & Te2 & $3a$ & (0, 0, 0)\\
		PdBi$_2$Te$_3$ & $R\bar{3}m$ & a = b = 4.421 & Pd & $3b$ & (0, 0, 0.5)\\
		& & c = 30.337 & Bi & $6c$ & (0, 0, 0.561)\\
		& & & Te1 & $3a$ & (0, 0, 0)\\
		& & & Te2 & $6c$ & (0, 0, 0.796)\\
		\hline
        \hline
	\end{tabular}
	\label{tab:convex-hull-struc}
\end{table*}

 \begin{table*}[!h]
 	\centering
 	\setlength{\tabcolsep}{5mm}
 	\caption{The total energy and formation energy of Pd-Bi-Te compounds used in the convex hull.}
 	\begin{tabular}{ccc}
 		\hline
         \hline
 		Compound & Energy(eV/f.u.) & Formation energy(eV/atom) \\
 		\hline
 		Pd & -5.161 & -- \\
 		Bi & -3.804 & -- \\
 		Te & -2.901 & -- \\
 		PdTe & -9.022 & -0.480 \\
 		PdTe$_2$ & -1.349 & -0.450 \\
 		PdBi & -9.565 & -0.3 \\
 		PdBi$_2$ & -13.568 & -0.266\\
 		BiTe & -7.346 & -0.320\\
 		Bi$_2$Te$_3$ & -18.258 & -0.389\\
 		Bi$_4$Te$_3$ & -25.979 & -0.294\\
 		PdBi$_2$Te$_3$ & -23.038 & -0.261\\
 		PdBi$_2$Te$_4$ & -26.849 & -0.354\\
 		PdBi$_2$Te$_5$ & -30.458 & -0.398\\
 		\hline
         \hline
 	\end{tabular}
 	\label{tab:convex-hull}
 \end{table*}

\section{The phonon spectrum, electronic structures and superconducting properties of PdBi$_2$Te$_4$}
\begin{figure}[t]
  \includegraphics[width=0.8\columnwidth]{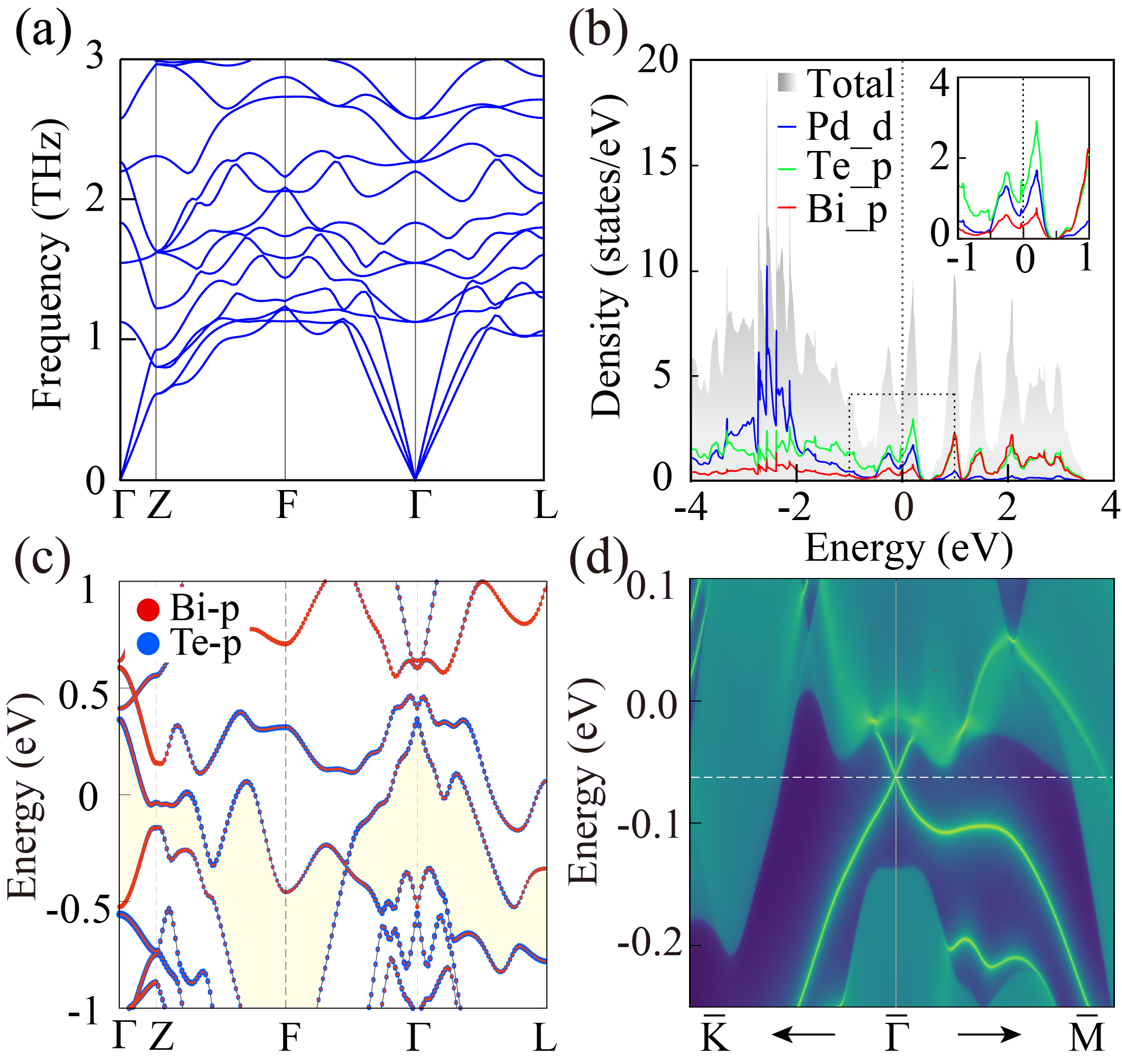}
  \caption{The phonon dispersion and electronic properties of PdBi$_2$Te$_4$.
  (a) The phonon dispersion of PdBi$_2$Te$_4$.
  (b) The total DOS and projected DOS of the Pd, Te, Bi atoms in PdBi$_2$Te$_4$,
  the zoom-in image shows the projected DOS near the Fermi level.
  (c) The orbital-projected band structures of PdBi$_2$Te$_4$, a continuous band gap around the Fermi level is marked by the yellow shade.
  (d) The topological surface states on (001) surface of PdBi$_2$Te$_4$.}
  \label{fig:S3}
\end{figure}

In Fig.~\ref{fig:S3}(a),
we calculate and plot the phonon dispersion of PdBi$_2$Te$_4$.
It clearly shows that there are 21 phonon modes with fully real positive frequencies,
indicating that the rhombohedral PdBi$_2$Te$_4$ are dynamically stable.
Based on the stable rhombohedral structure,
we carry out the calculations with SOC of electronic properties of PdBi$_2$Te$_4$.
The total and projected density of states (DOS) are shown in Fig.~\ref{fig:S3}(b),
which shows that the total DOS at the Fermi level is about of 3.95 states/eV,
indicating its metallic nature and the probability of superconductivity.
Fig.~\ref{fig:S3}(c) shows the projected band structures of PdBi$_2$Te$_4$, from which
we can see that two bands with $p$-orbital components from Te or Bi cross the Fermi level and
a continuous direct gap (the yellow region in Fig.~\ref{fig:S3}(c)) exists around the Fermi level.
Further detailed orbital components analysis demonstrates that
band inversion is present between the nominal valence band and conduction band,
indicating the non-trivial band topology of bulk PdBi$_2$Te$_4$.
In Fig.~\ref{fig:S3}(d),
we calculate and plot the topological surface states on the (001) surface of PdBi$_2$Te$_4$,
in which the Dirac surface states at the $\Gamma$ point manifest approximately $-62$~meV below the Fermi level,
confirming PdBi$_2$Te$_4$ is a $Z_2$ topological metal.

To investigate the superconducting properties of PdBi$_2$Te$_4$ and PdBi$_2$Te$_5$,
we calculate their electron-phonon coupling constant $\lambda$ and logarithmic average phonon frequency $\omega_{\log}$
as tabulated in Table.~\ref{tab:S5}.
As a comparison,
the $\lambda$ and $\omega_{\log}$ of parent SC PdTe and PdTe$_2$ are also calculated.
Moreover, the superconducting transition temperature ($T_c$)
is estimated by using the reduced Allen-Dynes formula as Eq.(4) in the main text.
By adopting a typical $\mu^*$ = 0.1, the $T_c$ of PdBi$_2$Te$_4$ and PdBi$_2$Te$_5$ is estimated as $3.11$~K and $0.57$~K, respectively.
Furthermore, the $T_c$ of PdTe and PdTe$_2$ is estimated as $2.55$~K and $1.59$~K,
which are consistent with the experimental results very well~\cite{Matthias1953, Karki2012, Leng2017,Das2018, Kudo2016}.
These results clearly demonstrate that the SC in PdTe and PdTe$_2$ is well inherited into the bulk of PdBi$_2$Te$_4$ and PdBi$_2$Te$_5$.

\begin{table*}[!h]
	\centering
	\setlength{\tabcolsep}{5.5mm}
	\caption{The calculated electron-phonon coupling strength $\lambda$, logarithmic average phonon frequency $\omega_{\log}$
	and the estimated $T_{c}$ ($\mu^*$ = 0.1) of PdTe, PdTe$_2$, PdBi$_{2}$Te$_{4}$ and PdBi$_{2}$Te$_{5}$.}
	\begin{tabular}{ccccc}
		\hline
        \hline
		Compound  &  $\lambda$  & $\omega_{\log}$(cm$^{-1}$) &  T$_{c}$(K)  &  T$_{c}^{exp}$(K)  \\
		\hline
		PdTe  &  0.61  & 106 & 2.55  &  2.3~\cite{Matthias1953}, 4.5~\cite{Karki2012} \\
		PdTe$_{2}$  &  0.52  & 112 & 1.59  &  1.64~\cite{Leng2017,Das2018,Kudo2016} \\
		PdBi$_{2}$Te$_{4}$  &  0.70 & 89 &  3.11  &  --  \\
		PdBi$_{2}$Te$_{5}$  &  0.43 & 97 &  0.57  &  -- \\
        \hline
        \hline
	\end{tabular}
	\label{tab:S5}
\end{table*}

\section{The chiral TSC phase in PdBi$_2$Te$_4$}
\begin{figure}[!h]
  \includegraphics[width=1.0\columnwidth]{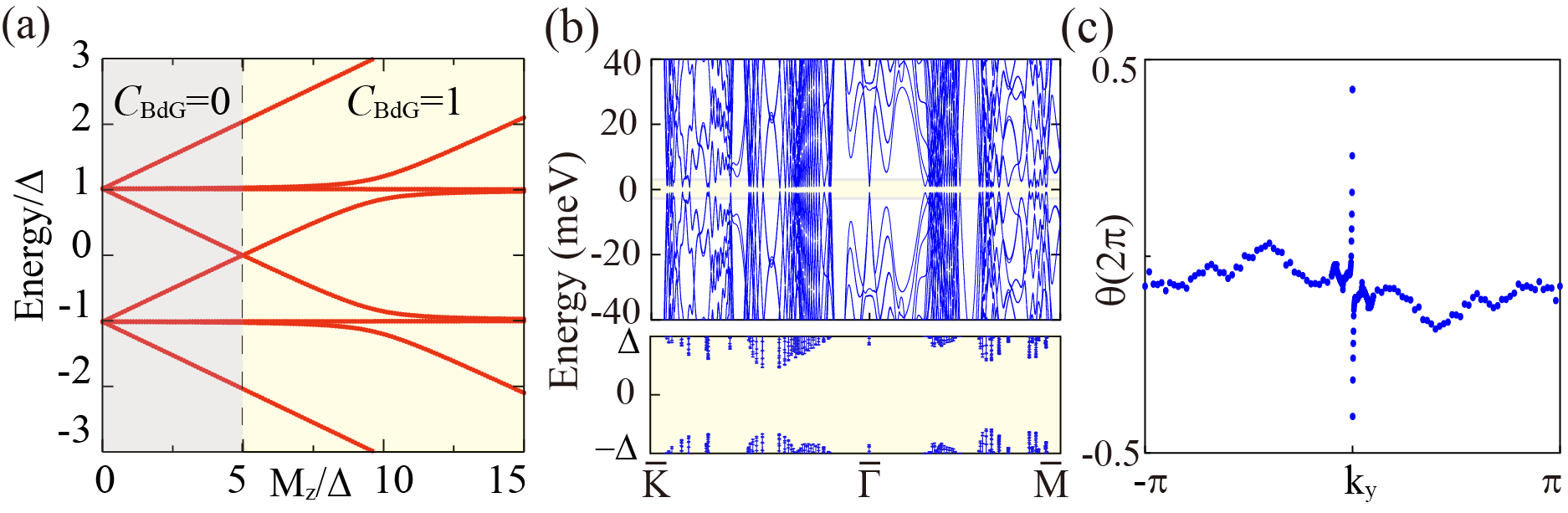}
  \caption{The chiral TSC phase in slab PdBi$_2$T$_4$.
  (a) The low energy spectrum at the $\Gamma$ point as the function of Zeeman splitting energy $M_z$, where $\Delta=1$~meV.
  (b) The superconducting spectrum along high symmetry paths with $M_z=10$~meV and $\Delta=1$~meV,
  the zoom-in image shows the full gap in the whole BZ.
  (c) The Wilson loop spectrum of all occupied states in (b),
  which manifest the superconducting BdG Chern number $C_{\textrm{BdG}}=1$ clearly.}
  \label{fig:S4}
\end{figure}

In this section, we study the chiral TSC phase in 20-SL PdBi$_2$Te$_4$ by incorporating its topological and superconducting properties with Zeeman splitting.
In this case,
the chemical potential $\mu$ is set at the energy
of surface Dirac cone at the $\Gamma$ point
(about $-62$ meV below the Fermi level),
the estimated $s$-wave superconducting gap $\Delta_s=1.0$~meV
is introduced globally for all 20-SLs,
and the out-of-plane Zeeman splitting is applied only in the bottom layer.
In Fig.~\ref{fig:S4}(a), we show the low energy spectrum of $H_{\mathrm{BdG}}$ at $\Gamma$
point as a function of Zeeman splitting energy $M_z$,
which manifests that two superconducting bands cross each other at critical point $M_z/\Delta=5$.
This behavior indicates that a topological phase transition happens at the critical point.
Fig.~\ref{fig:S4}(b) shows the superconducting energy dispersion with $M_z$=10 meV and $\Delta$=1 meV.
Its zoomed-in image, calculated in the whole BZ, directly reveals the full gap characters of superconducting spectrum,
indicating that the system is a well defined chiral TSC.
To further verify its nontrivial topological properties,
the superconducting BdG Chern number for all occupied states is calculated by the Wilson loop method as shown in Fig.~\ref{fig:S4}(c).
The spectrum of Wilson loop exhibits a non-trivial chiral winding number 1,
which directly confirms the superconducting BdG Chern number $C_{\textrm{BdG}}=1$.

\section{The electronic structures and topological properties of AuBi$_{2}$Te$_{5}$}
\begin{figure}[!h]
  \includegraphics[width=1.0\columnwidth]{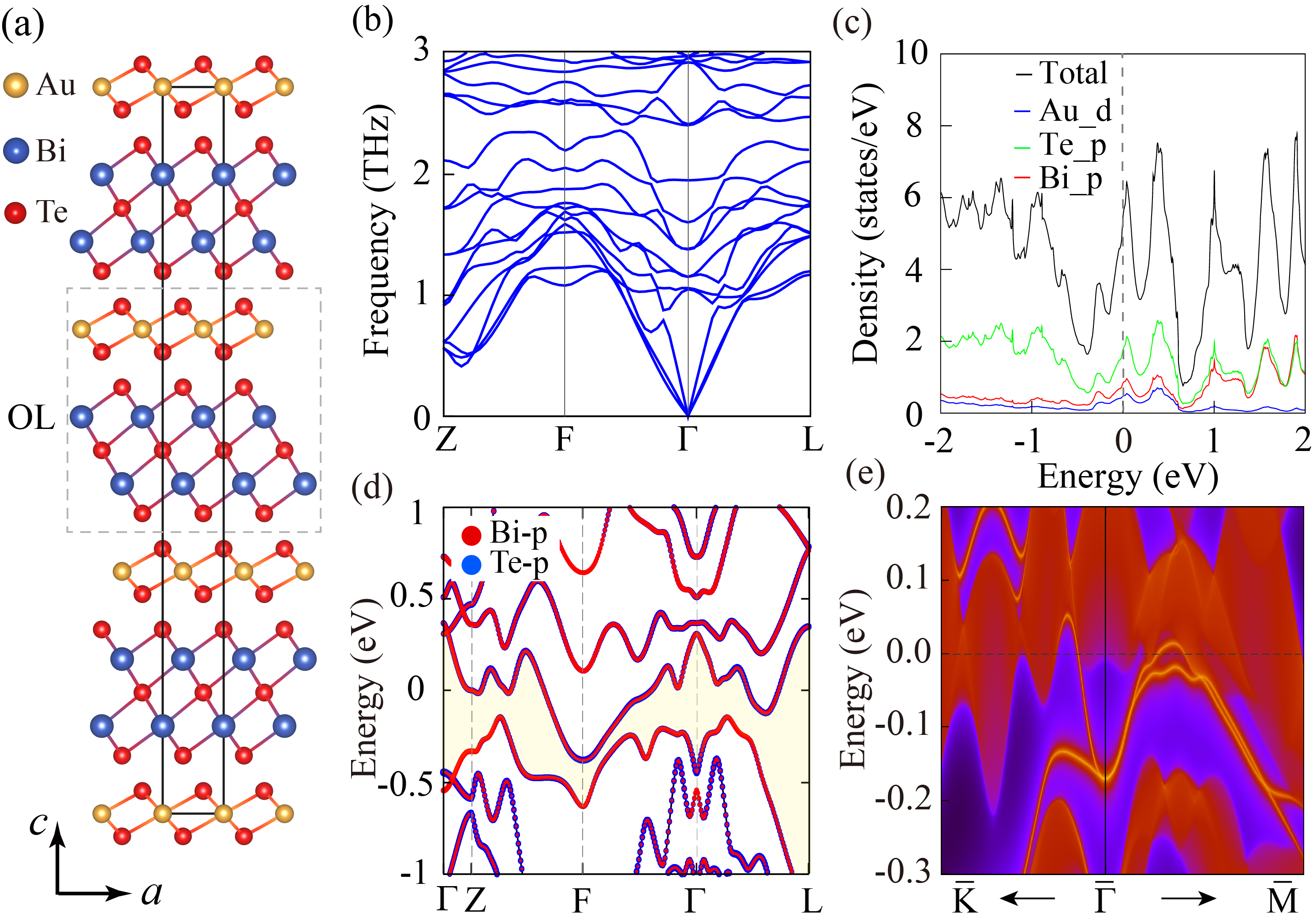}
  \caption{The crystal structure, phonon dispersion, and electronic properties of AuBi$_2$Te$_5$.
  (a) The side view of the crystal structure.
  (b) The phonon dispersion along high symmetry path.
  (d) The total DOS and projected DOS of the Au, Te, Bi atoms.
  (e) The orbital-projected band structures, a continuous band gap around the Fermi level is marked by the yellow shade.
  (f) The topological surface states on (001) surface.}
  \label{fig:S5}
\end{figure}

The crystal structure of rhombohedral AuBi$_2$Te$_5$ is shown in Fig.~\ref{fig:S5}(a),
which is formed by the layered intercalation of Bi$_2$Te$_3$ and AuTe$_2$.
In Fig.~\ref{fig:S5}(b), we show the calculated phonon dispersion
along the high symmetry lines of AuBi$_2$Te$_5$,
in which the most important observation
is that  all phonon modes have positive frequency throughout the BZ, indicating the dynamical stability of the rhombohedral structure.
In Fig.~\ref{fig:S5}(c), we calculate and plot the total
and projected DOS of AuBi$_2$Te$_5$, which gives rise to
DOS(0 eV)=5.41~states/eV at Fermi level,
indicating its metallic nature and the possibility of superconductivity.
Fig.~\ref{fig:S5}(d) shows the orbital-projected band structures of AuBi$_2$Te$_5$, in which a continuous band gap around the Fermi level is marked by the yellow shades.
Further detailed orbital components analysis demonstrates that a band inversion
exists between the nominal valence band and conduction band, which implies that AuBi$_2$Te$_5$ is a topological metal that inherits from parent Bi$_2$Te$_3$.
In Fig.~\ref{fig:S5}(e), we show the calculated surface states of AuBi$_2$Te$_5$ in the (001) surface.
It clearly shows that the topological surface states are presented in the bulk gap, confirming AuBi$_2$Te$_5$ is a $Z_2$ topological metal.

